\documentclass[usenatbib]{mnras}

\usepackage{lscape}
\usepackage{graphicx}
\usepackage{amssymb}
\usepackage{amsmath}
\usepackage{url}
\usepackage{bm}
\usepackage{aas_macros}
\usepackage{rotating}

\usepackage{multirow}

\title[Intrinsic reddening of the Magellanic Clouds]{The intrinsic reddening of the Magellanic Clouds as traced by background galaxies -- I.
The bar and outskirts of the Small Magellanic Cloud}

\author[C.~P.~M.~Bell et al.]{Cameron~P.~M.~Bell,${^1}$\thanks{E-mail:
  cbell@aip.de (CPMB)} Maria-Rosa L. Cioni,$^{1}$ A.~H.~Wright,$^{2}$ Stefano Rubele,$^{3,4}$
  \newauthor David~L.~Nidever,$^{5,6}$ Ben~L.~Tatton,$^{7}$ Jacco Th. van Loon,$^{7}$ Valentin D. Ivanov,$^{8,9}$
  \newauthor Smitha Subramanian,$^{10}$ Joana M. Oliveira,$^{7}$ Richard de Grijs,$^{11,12,13}$
  \newauthor Clara M. Pennock,$^{7}$ Yumi Choi,$^{5,14}$ Dennis Zaritsky,$^{14}$ Knut Olsen,$^{6}$
  \newauthor Florian Niederhofer,$^{1}$ Samyaday Choudhury,$^{11,12}$ David Mart{\'i}nez-Delgado,$^{15}$
  \newauthor Ricardo R. Mu{\~n}oz$^{16}$\\
  $^{1}$Leibniz-Institut f{\"u}r Astrophysik Potsdam (AIP), An der Sternwarte 16, D-14482 Potsdam, Germany\\
  $^{2}$Argelander-Institut f{\"u}r Astronomie, Universit{\"a}t Bonn, Auf dem H{\"u}gel 71, D-53121 Bonn, Germany\\
  $^{3}$Dipartimento di Fisica e Astronomia, Universit{\`a} di Padova, Vicolo dell'Osservatorio 2, I-35122 Padova, Italy\\
  $^{4}$Osservatorio Astronomico di Padova -- INAF, Vicolo dell'Osservatorio 5, I-35122 Padova, Italy\\
  $^{5}$Department of Physics, Montana State University, P.O. Box 173840, Bozeman, MT 59717, USA\\
  $^{6}$National Optical Astronomy Observatory, 950 North Cherry Avenue, Tucson, AZ 85719, USA\\
  $^{7}$Lennard-Jones Laboratories, Keele University, ST5 5BG, UK\\
  $^{8}$European Southern Observatory, Ave. Alonso de C{\'o}rdova 3107, Vitacura, Santiago, Chile\\
  $^{9}$European Southern Observatory, Karl-Schwarzschild-Str. 2, 85748 Garching bei MŸnchen, Germany\\
  $^{10}$Indian Institute of Astrophysics, Koramangala II Block, Bangalore, 560034, India\\
  $^{11}$Department of Physics and Astronomy, Macquarie University, Balaclava Road, Sydney NSW 2109, Australia\\
  $^{12}$Research Centre for Astronomy, Astrophysics and Astrophotonics, Macquarie University, Balaclava Road, Sydney NSW 2109, Australia\\
  $^{13}$International Space Science Institute--Beijing, 1 Nanertiao, Zhongguancun, Hai Dian District, Beijing 100190, China\\
  $^{14}$Steward Observatory, University of Arizona, 933 North Cherry Avenue, Tucson, AZ 85721, USA\\
  $^{15}$Astronomisches Rechen-Institut, Zentrum f{\"u}r Astronomie der Universit{\"a}t Heidelberg, M{\"o}nchhofstr. 12-14, 69120, Heidelberg, Germany\\
  $^{16}$Departamento de Astronom{\'i}a, Universidad de Chile, Camino del Observatorio 1515, Las Condes, Santiago, Chile}
\begin{document}

\date{Accepted ?, Received ?; in original form ?}

\pagerange{\pageref{firstpage}--\pageref{lastpage}} \pubyear{2019}

\maketitle

\label{firstpage}

\begin{abstract}
We present a method to map the total intrinsic reddening of a foreground extinguishing medium via the analysis of spectral energy distributions (SEDs) of background
galaxies. In this pilot study, we implement this technique in two distinct regions of the Small Magellanic Cloud (SMC) -- the bar and the southern outskirts -- using a
combination of optical and near-infrared $ugrizYJK_{\mathrm{s}}$ broadband imaging. We adopt the \textsc{lephare} $\chi^{2}$-minimisation SED-fitting routine and
various samples of galaxies and/or quasi-stellar objects to investigate the intrinsic reddening. We find that only 
when we construct reddening maps using objects classified as galaxies with low
levels of intrinsic reddening (i.e. ellipticals/lenticulars and early-type spirals), the resultant maps are consistent with previous literature determinations i.e. the intrinsic
reddening of the SMC bar is higher than that in the outer environs.
We employ two sets of galaxy templates -- one theoretical and one empirical -- to test for template dependencies
in the resulting reddening maps and find that the theoretical templates imply systematically higher reddening values by up to 0.20\,mag in $E(B-V)$.
A comparison with previous reddening maps, based on the stellar components of the SMC, typically shows reasonable agreement. There is, however, significant
variation amongst the literature reddening maps as to the level of intrinsic reddening associated with the bar. Thus, it is difficult to unambiguously state that instances of
significant discrepancies are the result of appreciable levels of dust not accounted for in some literature reddening maps or whether they reflect issues with our
adopted methodology.
\end{abstract}


\section{Introduction}
\label{introduction}

The Large and Small Magellanic Clouds (LMC and SMC)
are the largest, most massive dwarf satellite galaxies of the Milky Way (MW), and
represent an unparallelled testbed to study the formation
and evolution of galaxies. Given the interactions between the LMC and SMC
as well as the Magellanic Clouds (MCs) and the MW
(see e.g \citealp{Olsen11,Besla12,Diaz12,Choi18}), they provide a unique insight
into the effects of such interactions on the structure and history of the
galaxies. In a cosmological context, interacting galaxies can also be used
to study the effects of minor merger events, which are believed to be important
for the formation of MW-like galaxy haloes \citep{Brook04,Read08}.

Studying interacting galaxies, however, is non-trivial
as the underlying structure and stellar populations have
both been shaped by prior events. A dominant factor contributing
to the uncertainties is the internal dust of the galaxy, which is generally
non-uniform and can thus have a significant effect on the derived physical properties.
The reddening is typically inferred through colour-magnitude diagram (CMD) analyses.

The reddening of the MCs has been extensively investigated, based predominantly on various
stellar components of the MCs. Such methods include comparing the colour of the red clump
and/or RR Lyrae stars with the unreddened colour predicted by stellar evolutionary models
at the corresponding metallicity (see e.g. \citealp{Udalski99a,Udalski99b,Subramanian09,Subramanian12}; \citealp*{Haschke11}; \citealp{Tatton13};
\citealp{Choi18}), multi-band fitting of the reddening law to the apparent distance
moduli of Cepheids to derive individual reddening values (see e.g. \citealp{Inno16}),
as well as comparing multi-band photometry with stellar atmospheric models
(see e.g. \citealp{Zaritsky02,Zaritsky04}). In addition, longer-wavelength mid-/far-infrared (IR), microwave and
radio observations of the MCs have also played crucial roles in not only mapping, but also quantifying the dust
content of the MCs (see e.g. \citealp{Gordon03,Leroy07,Bernard08,Israel10}).

Here we present a pilot study to map and quantify the total intrinsic reddening of the MCs based on observations
of background galaxies behind the MCs. Whilst background galaxies have been used in the past to probe the reddening
of the MCs (see e.g. \citealp{Hodge69,Gurwell90,vanLoon97,Dutra01,Krienke01}), these studies have typically used only a small number of galaxies
to place constraints on the reddening along given line of sights based on either their observed colours or through spectroscopic analysis.
In this study, we combine optical and near-IR photometric data to construct eight-band spectral energy distributions (SEDs) which we use to
map the reddening across the MCs as well as determine the total line-of-sight reddening. This technique has the advantage
that it does not rely on the use of stars in the CMD i.e. it is
independent of the tool used to study the underlying stellar populations. Furthermore,
as it does not involve the use of variable stars, the technique is not spatially limited to those
which have been catalogued through time-series photometric identification of light curves.

The structure of the paper is as follows. In Section~\ref{photometric_dataset} we introduce the photometric data sets,
describe our galaxy selection procedure and also discuss the photometric pipeline used to create the SEDs. Section~\ref{fitting_seds_of_galaxies}
describes the fitting of the galaxy SEDs and 
in Section~\ref{determining_intrinsic_reddening_magellanic_clouds} we introduce the method adopted to determine the intrinsic reddening.
We discuss our results in Section~\ref{discussion} and in Section~\ref{summary} we present our conclusions.

\section{Photometric data set}
\label{photometric_dataset}

Our data set consists of optical and near-IR photometry taken as part of
the Survey of the MAgellanic Stellar History (SMASH; \citealp{Nidever17}) and the
Visible and Infrared Survey Telescope for Astronomy (VISTA)
survey of the Magellanic Clouds system (VMC; \citealp{Cioni11}), which together are composed of
eight bandpasses ($u, g, r, i, z, Y, J, K_{\rm{s}}$) covering the wavelength range 0.3--2.5\,$\mu$m.

\subsection{SMASH optical data}
\label{smash_data}

SMASH is a National Optical Astronomy Observatory (NOAO) community Dark Energy Camera (DECam) survey
of the MC system which maps $\sim$\,480\,deg$^{2}$ (distributed across an area of $\sim$\,2400\,deg$^{2}$ at a 20 per cent
filling factor) with limiting (AB) magnitudes of $\sim$\,24\,mag in $ugriz$.
DECam \citep{Flaugher15} mounted on the Blanco 4-m telescope at Cerro Tololo Inter-American
Observatory (CTIO) comprises 62 CCDs (of which $59\frac{1}{2}$ were functional throughout the majority
of the SMASH observations; see \citealp{Nidever17} for details) covering a field of view of 3\,deg$^{2}$ with
a pixel scale of 0.27$''$. All scheduled SMASH observations have now been completed from which we
determine a median image seeing of 1.21, 1.12, 1.00, 0.94 and
0.89\,arcsec in $ugriz$, respectively, with a standard deviation ranging from 0.32 ($u$) to 0.24\,arcsec ($z$). 
To cover the gaps between the CCDs, the survey's observing strategy consists
of deep exposures (999\,s in $uiz$ and 801\,s in $gr$), in addition to three sets of short 60\,s exposures in $ugriz$ with
large, half-CCD offsets between each set.

\subsection{VMC near-IR data}
\label{vmc_data}

The VMC is one of the first six public surveys commissioned by the European Southern Observatory (ESO) and covers
$\sim$170\,deg$^{2}$ of the MC system with limiting (AB) magnitudes of $\sim$\,22\,mag in $YJK_{\rm{s}}$.
The VISTA IR CAMera (VIRCAM; \citealp{Sutherland15}) comprises 16 detectors arranged in a sparse-filled 4\,$\times$\,4 rectangular
grid within the 1.65 degree diameter field of view, with a pixel scale of 0.34$''$, and provides an active field of 0.60\,deg$^{2}$ on
pixels (known as a \emph{pawprint}). Due to gaps of 0.9 detector width in the $x$-direction and 0.425 detector width in the
$y$-direction a set of six offset pawprints are observed to give one filled rectangular area consisting of a central rectangle of
1.475\,$\times$\,1.017 degrees covered by at least two of the six pawprints (1.5\,deg$^{2}$); plus two thin stripes each
0.092\,degrees wide (along the two long edges) covered by just one pawprint, giving a total contiguously covered area of
1.77\,deg$^{2}$ (known as a \emph{tile}).
VMC observations were completed in October 2018 and from these we determine a median image seeing
of 1.03, 1.00 and 0.93\,arcsec in $YJK_{\rm{s}}$ with standard deviations of 0.13, 0.10 and 0.08\,arcsec, respectively.
The VMC is a multi-epoch survey providing three epochs in $YJ$ and 12 in $K_{\rm{s}}$, so accounting for the
individual pawprint exposures, the number of offset exposures and the number of epochs in each band, the
total exposure times per tile are 2400\,s in $YJ$ and 9000\,s in $K_{\rm{s}}$.
Repeat observations within each survey suggest a photometric precision of better than 2 per cent in all bands.

\begin{figure}
\centering
\includegraphics[width=\columnwidth]{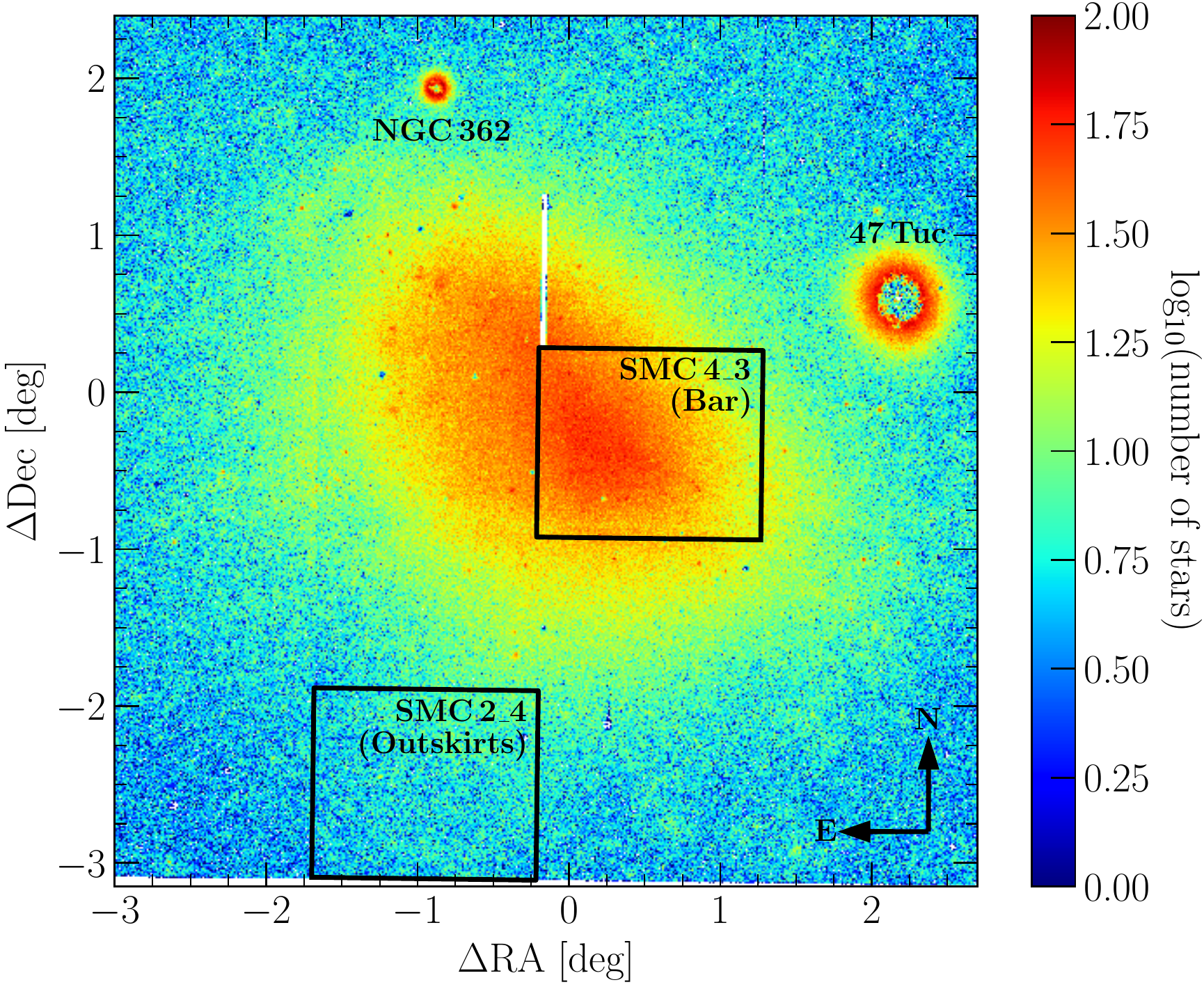}
\caption[]{Stellar density plot of the central region of the SMC as observed by the VMC survey. The origin of the plot corresponds to the
centre of the SMC as given by the HYPERLEDA catalogue of \cite{Paturel03}. The black rectangles denote the positions of the two tiles focussed
on in this pilot study. The positions of the prominent foreground Milky Way globular clusters 47\,Tuc and NGC\,362 are also shown.
The underdensities in the cores of both clusters are due to completeness effects resulting from crowding in the central regions.
The vertical stripe at $\Delta$RA\,$\simeq -0.2$\,deg is due to a narrow gap in the VMC observations.}
\label{fig:star_dens_tiles}
\end{figure}

This pilot study represents a proof of concept for our technique to determine the total intrinsic reddening using
galaxies observed through the MCs and as such we focus on only two VMC tiles.
The complete analysis of the combined SMASH-VMC
footprint is still underway. In this paper we focus on tiles SMC\,2\_4 and SMC\,4\_3 from the VMC survey -- hereafter
referred to as the outskirts and bar regions of the SMC -- centred at
$(\alpha, \delta)_{\mathrm{J2000.0}} = 01^{\mathrm{h}}\,07^{\mathrm{m}}\,33.9^{\mathrm{s}},
-75^{\circ}\,15'\,59.8''$ and $00^{\mathrm{h}}\,45^{\mathrm{m}}\,14.7^{\mathrm{s}},
-73^{\circ}\,07'\,11.3''$, respectively. Fig.~\ref{fig:star_dens_tiles} shows the positions of these tiles
overlaid on a stellar density plot of the central region of the SMC for which we have adopted the
centre of the SMC ($00^{\mathrm{h}}\,52^{\mathrm{m}}\,38.0^{\mathrm{s}}, -72^{\circ}\,48'\,01.0''$) as provided
by the HYPERLEDA catalogue of \cite{Paturel03}. This figure demonstrates that
the two tiles probe regions of different stellar density of the SMC, and sample large enough areas to encompass different levels of intrinsic reddening.

\subsection{\textit{A priori} identification of background galaxies}
\label{identification_of_galaxies}

To identify background galaxies we adopt a combination of colour-magnitude and morphological
selections. Given the relatively crowded nature of the MCs and the cleaner
separation between stars and galaxies in near-IR CMDs (see e.g. \citealp{Maddox12,Cioni14}), compared to
their optical counterparts (see e.g. \citealp{Nidever19}), we opt to identify background galaxies
using the VMC PSF photometric catalogues. These
catalogues are generated from homogenised deep tiles which are created by combining pawprints
from different epochs (see \citealp{Rubele15}),
which ``smears'' any variability.
Pawprints which were taken in sub-optimal observing conditions (e.g. high seeing,
patchy cloud cover, etc.) were not included in the creation of the deep tiles.

\begin{figure}
\centering
\includegraphics[width=\columnwidth]{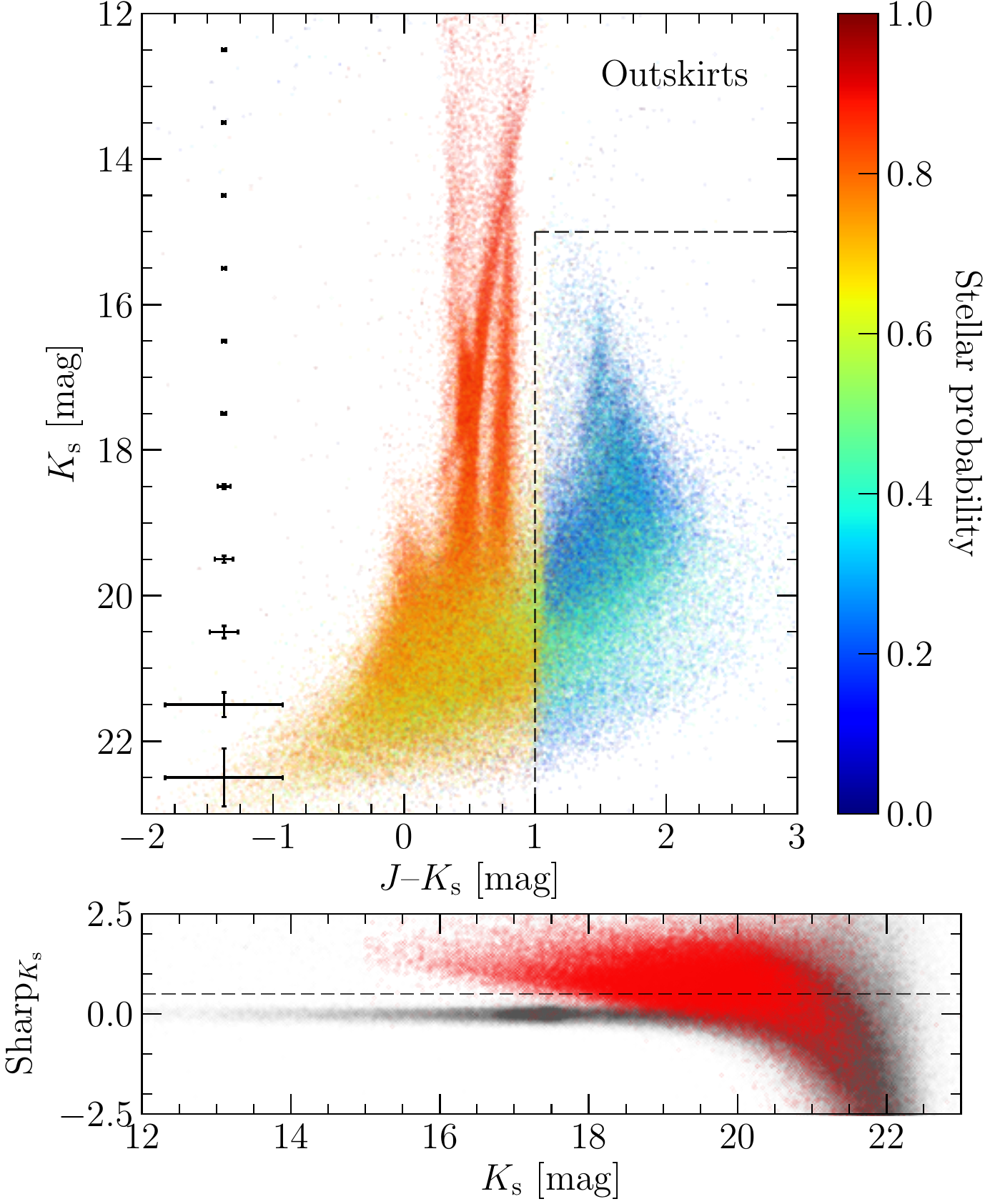}
\caption[]{\textit{Top panel:} $K_{\rm{s}}$, $J$--$K_{\rm{s}}$ CMD of the outskirts region of the SMC colour-coded
as a function of stellar probability. The dashed lines denote our initial colour-magnitude selection
($J$--$K_{\rm{s}} > 1.0$\,mag and $K_{\rm{s}} > 15$\,mag) used to identify galaxies. The error bars on the left
denote the median photometric errors as a function of $K_{\mathrm{s}}$-band magnitude at 1\,mag intervals.
\textit{Bottom panel:} $K_{\rm{s}}$-band
sharpness index as a function of $K_{\rm{s}}$-band magnitude. The black points correspond to
the points shown in the top panel, whereas the red points denote the subset of objects which satisfied our initial
colour-magnitude selection. The horizontal dashed line denotes our adopted $K_{\rm{s}}$-band sharpness index
threshold (Sharp$_{K_{\rm{s}}} > 0.5$) to refine our galaxy selection.}
\label{fig:cmd_sharp}
\end{figure}

We initially define our sample of galaxies as those objects which satisfy both $J$--$K_{\rm{s}} > 1.0$\,mag and
$K_{\rm{s}} > 15$\,mag in the $K_{\rm{s}}$, $J$--$K_{\rm{s}}$ CMD, thereby ensuring
that our objects are detected in at least two bands. We impose the bright magnitude cut to avoid including
intrinsically reddened evolved stars and/or young stellar objects in our sample.
The top panel of Fig.~\ref{fig:cmd_sharp} shows our initial
colour-magnitude selection for the outskirts region of the SMC. This sample is further refined using a combination of the
stellar probability and $K_{\rm{s}}$-band sharpness index\footnote{The \textsc{daophot} sharpness index
is proportional to the ratio of the FWHM of the object to the FWHM of the PSF model and so extended
objects will tend to have values appreciably larger than zero.} provided for each unique object as part of the
VMC PSF catalogues. Stellar probabilities are calculated based on a combination of the position
in the colour-colour diagram, the local completeness level (as determined by artificial star tests; see
\citealp{Rubele11}) and the \textsc{daophot} \citep{Stetson87} sharpness index.
Fig.~\ref{fig:cmd_sharp} clearly shows that the objects initially identified through our
colour-magnitude selection have low stellar probabilities (mean value of 0.33) and also exhibit significantly larger
sharpness indices than one would expect for stellar objects. We adopt the sharpness index as measured
from the $K_{\rm{s}}$-band images as the VMC observing requirements ensure the highest quality seeing in this
bandpass (see \citealp{Cioni11}). We therefore include the criteria that objects which satisfy our initial
colour-magnitude selection must also have stellar probabilities of less than 0.33 and $K_{\rm{s}}$-band
sharpness indices of greater than 0.5. These combined criteria result in the identification of 29,843 and
18,509 potential galaxies in the outskirts and bar regions of the SMC, respectively.

There will of course be a number of sources with stellar probabilities of less than 0.33 at bluer colours. The available
colour-magnitude and morphological information is insufficient to differentiate between whether such objects are
actually galaxies or simply blended stars, and so to retain a \emph{relatively} clean sample of galaxies we do
not loosen the aforementioned criteria. The objects which meet these selection criteria, however, will invariably include
a number of both quasi-stellar objects (QSOs) and stars (especially at fainter magnitudes
where the $K_{\rm{s}}$-band sharpness index distribution begins to bloom; see the bottom panel of Fig.~\ref{fig:cmd_sharp}).
Of these, stars are genuine contaminants as they are not observed through the MCs and hence their SEDs do not include the imprints of the
intrinsic MC reddening. QSOs, on the other hand, can be used in exactly the same way as the background galaxies. Further details pertaining to
how we differentiate between different types of objects (galaxies, QSOs and stars) are presented in Section~\ref{fitting_seds_of_galaxies} in which we describe
our adopted SED-fitting code.

\subsection{{\sc{lambdar}} photometry}
\label{lambdar_photometry}

Fluxes for each of our targets are measured using 
the Lambda Adaptive Multi-Band Deblending Algorithm in R ({\sc lambdar}, v0.20; 
\citealt{Wright16}). \textsc{lambdar} is designed to perform accurate matched 
aperture photometry, of pre-defined sources, across images that are neither 
pixel nor PSF matched. To perform the photometry \textsc{lambdar} requires two inputs;
a catalogue of objects containing positions and aperture parameters as well as a FITS image with World
Coordinate System (WCS) astrometry. We also took advantage of the \textsc{lambdar} functionality to mask out
contaminants to remove the contribution of sources within 3\,arcsec of each object.

\begin{figure}
\centering
\includegraphics[width=\columnwidth]{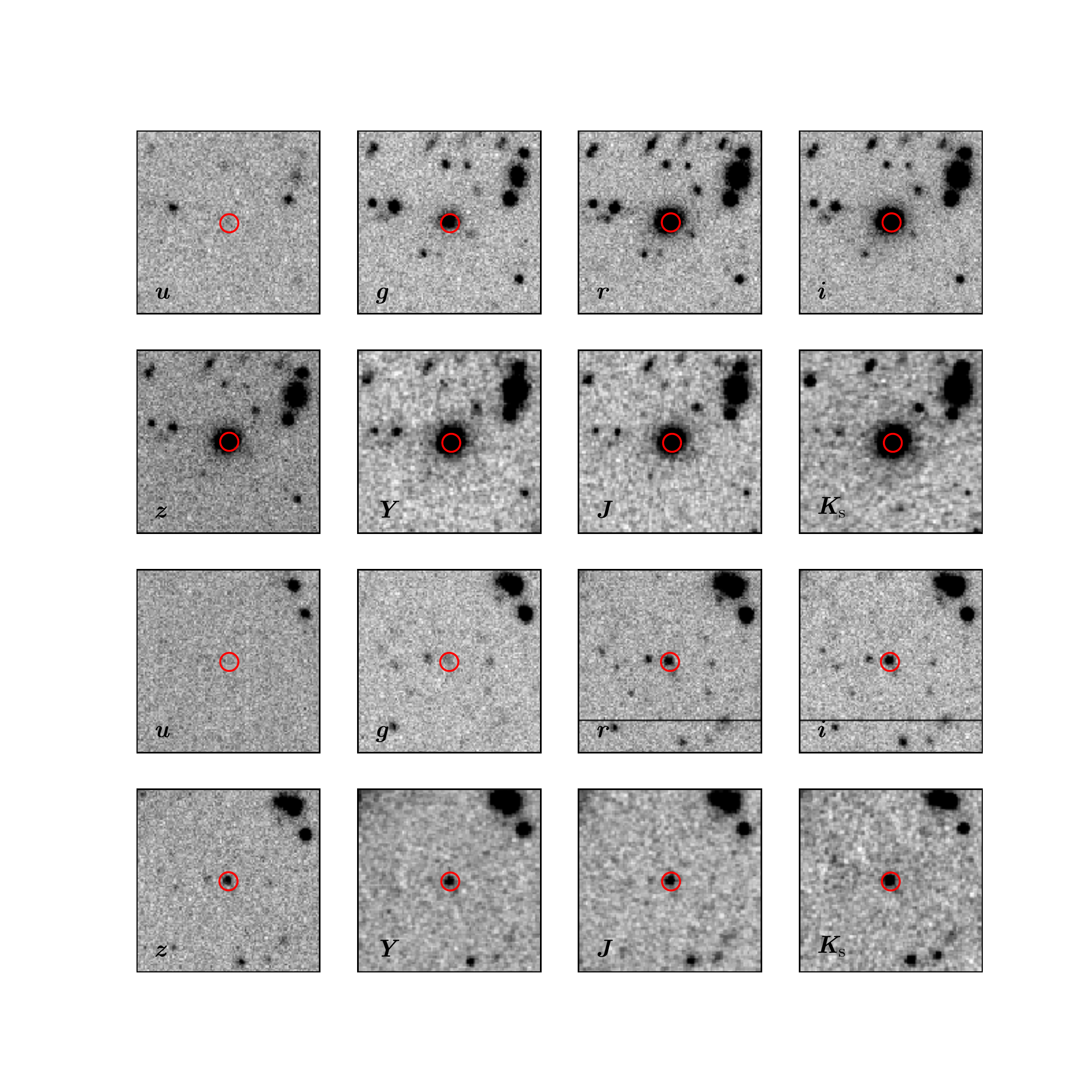}
\caption[]{Postage stamp cut-outs of typical $K_{\rm{s}}$\,$\simeq$\,18 and 20\,mag galaxies (top two and bottom two rows,
respectively) in the outskirts region of the SMC in each of the eight bandpasses.
Each image is 0.5$\,\times$\,0.5\,arcmin and orientated such that
north is up and east is to the left. The red circle in each panel denotes the adopted 3\,arcsec diameter aperture
used to measure the fluxes for the standard aperture photometric catalogues (see the text for details).}
\label{fig:postage_stamps}
\end{figure}

As a result of crowding, we use \textsc{lambdar} to not only measure fluxes using standard (circular) apertures
of diameter 3\,arcsec [see e.g. the Cosmic Evolution Survey (COSMOS; \citealp{Scoville07}) optical and near-IR
photometric catalogue produced by \citealp{Capak07}], but also PSF-weighted apertures which will result in more
reliable deblending of sources. PSF-weighted apertures
preferentially extract information from the central regions of galaxies which acts to both reduce the noise
on the estimates caused by noisy low-flux pixels in the galaxy's outskirts and reduce the complexity of the SED
required to model the galaxy by probing the (typically redder) central regions of the galaxy. By using fluxes extracted using PSF-weighted apertures, however, one implicitly
ignores the self-contamination of the inner-galaxy profile by flux from the outskirts of the galaxy. If the PSF between different
bands varies significantly, this could result in slightly incorrect flux estimates and/or systematics, however as the size of the PSFs
across all eight bands are similar (see Sections~\ref{smash_data} and \ref{vmc_data}), this essentially means that the contamination is consistent in
each band and thus should not significantly affect the resultant SED. Fig.~\ref{fig:postage_stamps} shows
postage stamp cut-outs of typical $K_{\rm{s}} \simeq 18$ and 20\,mag galaxies in the outskirts region of the SMC in each of the eight
bandpasses with a 3\,arcsec diameter aperture overlaid.

Specific details of \textsc{lambdar} are presented by \cite{Wright16}, although 
there have been some developments to the code since its publication which are 
relevant to the work presented here. In particular, the program now includes the 
functionality to estimate the PSF of an input image by stacking suitably segregated 
high-signal-to-noise point sources. As the imaging under analysis here can be significantly
affected by crowding, this functionality is able to return high fidelity PSFs and thus is 
particularly useful. To ensure that our final PSF estimate is robust, we have generated PSFs
for a given image using different samples of point sources. 
The PSF estimates are seen to vary by less than half a percent in integrated area, suggesting that
the choice of which point sources contribute to the stacked PSF is not critical to the fidelity 
of the PSF estimate.
Other than the new PSF estimation, the code remains largely unchanged since v0.12 described by 
\cite{Wright16}. Nonetheless, some minor changes and genuine bug fixes have been implemented, and 
are all documented in the GitHub repository commit history\footnote{\url{https://github.com/AngusWright/LAMBDAR}}.

For our input FITS images we downloaded the VMC images covering the bar and outskirts regions of the SMC from the 
VISTA Science Archive\footnote{\url{http://horus.roe.ac.uk/vsa}. VMC data for the bar region of the SMC (SMC\,4\_3) are publicly available and those
for the outskirts (SMC\,2\_4) will soon be released as part of VMC Data Release 5 which is in preparation.} (VSA; \citealp{Cross12}).
We adopt the \emph{deep stacks} (the stacked pawprints from different epochs, each of which comprises 16 detectors) as opposed to
the single \emph{deep tiles} as the latter can suffer from significant jumps in the sky background level on small
scales due to the process of combining different pawprints in the creation of the tile\footnote{Despite known issues with some of the
VIRCAM detectors (see \url{http://casu.ast.cam.ac.uk/surveys-projects/vista/technical/known-issues} for a full description), we do not notice
any obvious trends in the reddening maps which would suggest that these significantly affect our results.}.
To calibrate the near-IR data onto an AB magnitude system, we follow the prescription of
\citet[see specifically the offsets listed in eqns.~19--22 and D2--D6]{Gonzalez-Fernandez18}.

To calibrate the optical data onto a native DECam AB magnitude system we remove the SMASH colour terms introduced to
calibrate the photometry onto a quasi-SDSS system. This process was performed on a CCD-by-CCD basis
and involves the use of ``fiducial'' colours (as defined by the SDSS standard stars and which includes the band one wishes to calibrate;
see \citealp{Nidever17}).
For our ``fiducial'' colours we adopt the following values: $u$--$g = 1.360$, $g$--$r = 0.667$, $g$--$r = 0.586$, $i$--$z = 0.318$,
$i$--$z = 0.347$\,mag for the $u$-, $g$-, $r$-, $i$- and $z$-bands, respectively. The uncertainties on the coefficients used in the SMASH
transformation equations are very small (less than 1 per cent) and as we perform a transformation with a fixed fiducial colour the resultant
uncertainties on the DECam AB magnitude photometry should be equally small.

\begin{table}
\caption[]{Number of objects in each of the samples.
The values in parentheses denote the percentage of the total number of objects for which positive fluxes
have been calculated in only eight,
seven, six, five and four bands, respectively.}
\begin{tabular}{l c c}
\hline
Region   &   Sample   &   No. of objects\\
\hline
\multirow{2}{*}{Outskirts}   &   Aper.   &   29,606 (91.4, 6.1, 1.5, 0.5, 0.5)\\
   &   PSF   &   29,788 (93.9, 4.7, 0.9, 0.3, 0.2)\\
\multirow{2}{*}{Bar}   &   Aper.   &   16,796 (66.3, 18.9, 8.0, 4.2, 2.6)\\
   &   PSF   &   18,248 (74.2, 15.9, 6.2, 2.6, 1.1)\\
\hline
\end{tabular}
\label{tab:catalogue_properties}
\end{table}


Given the observing strategy of both the SMASH and VMC surveys, each galaxy in our sample is observed
multiple times except for objects in two narrow strips along the length of the VIRCAM array (see fig.~3 of
\citealp{Cross12}). Although these regions, except for the lower strip of the outskirts region of the SMC which corresponds to the
lower limit of the VMC coverage, are covered by observations of other tiles in the VMC survey, we
do not include these additional measurements. The final flux for each object is simply the weighted mean of all available
measurements in that band and for which we set the weights equal to the inverse square of the corresponding uncertainty
on the flux. To ensure reliable SED fits, we only retain objects for which we measure positive fluxes in at least four out of the eight
available bandpasses. Given that we have extracted fluxes
using both standard (circular) as well as PSF-weighted apertures, the number of resultant objects for a given tile in both
cases is not necessarily the same. Our final standard aperture and PSF-weighted aperture photometric catalogues have a total
of 29,606 and 29,788 objects, respectively, for the outskirts region of the SMC, whereas for the bar region the catalogues contain
16,796 and 18,248 objects (see Table~\ref{tab:catalogue_properties}).

\subsection{Accounting for foreground Milky Way reddening}
\label{foreground_reddening}

The final step in the creation of our galaxy SEDs is to account for diffuse foreground MW reddening.
The line-of-sight reddening towards a given galaxy includes several components, and thus by removing the
MW component this will make it easier to infer the intrinsic reddening of the two SMC tiles from the SEDs.
We determine the corresponding $E(B-V)$ value for a given
position using the \texttt{Python} module \textsc{sfdmap}\footnote{\url{https://github.com/kbarbary/sfdmap}}
which linearly interpolates within the \citet*[hereafter SFD98]{Schlegel98} dust maps. Following \cite{Schlafly11}, a scaling
of 0.86 is applied as standard to the SFD98 $E(B-V)$ values to reflect the fact that these values were
calculated using the \cite*{Cardelli89} and \cite{ODonnell94} extinction laws, which have since been shown
to overestimate the resultant reddening values (see also \citealp{Schlafly10}; \citealp*{Yuan13}). To convert a
given $E(B-V)$ value to extinction in a specific bandpass, we adopt the following coefficients:

\begin{equation}
\frac{A_{\lambda}}{E(B-V)} = \left\{
  \begin{array}{ll}
    3.963; & \lambda = u \\
    3.186; & \lambda = g \\
    2.140; & \lambda = r \\
    1.569; & \lambda = i \\
    1.196; & \lambda = z \\
    1.017; & \lambda = Y \\
    0.705; & \lambda = J \\
    0.308; & \lambda = K_{\mathrm{s}}.
    \end{array}
  \right.
\label{eqn:ext_coeffs}
\end{equation}

\noindent The $griz$ coefficients are taken from the Dark Energy Survey (DES) Data Release (DR) 1
paper by \cite{Abbott18}, whereas the $YJK_{\rm{s}}$ coefficients are from
\cite{Gonzalez-Fernandez18}\footnote{As the DES does not include $u$-band observations
the corresponding coefficient is not given by \cite{Abbott18}. This was instead calculated in exactly the same way as for
the $griz$ coefficients and kindly provided upon request by the DES collaboration.}.
These coefficients have been calculated assuming $R_{V}=3.1$
as well as account for the \cite{Schlafly11} recalibration, and should therefore be used in conjunction with
the unscaled SFD98 $E(B-V)$ values provided by \textsc{sfdmap}.

\begin{figure}
\centering
\includegraphics[width=\columnwidth]{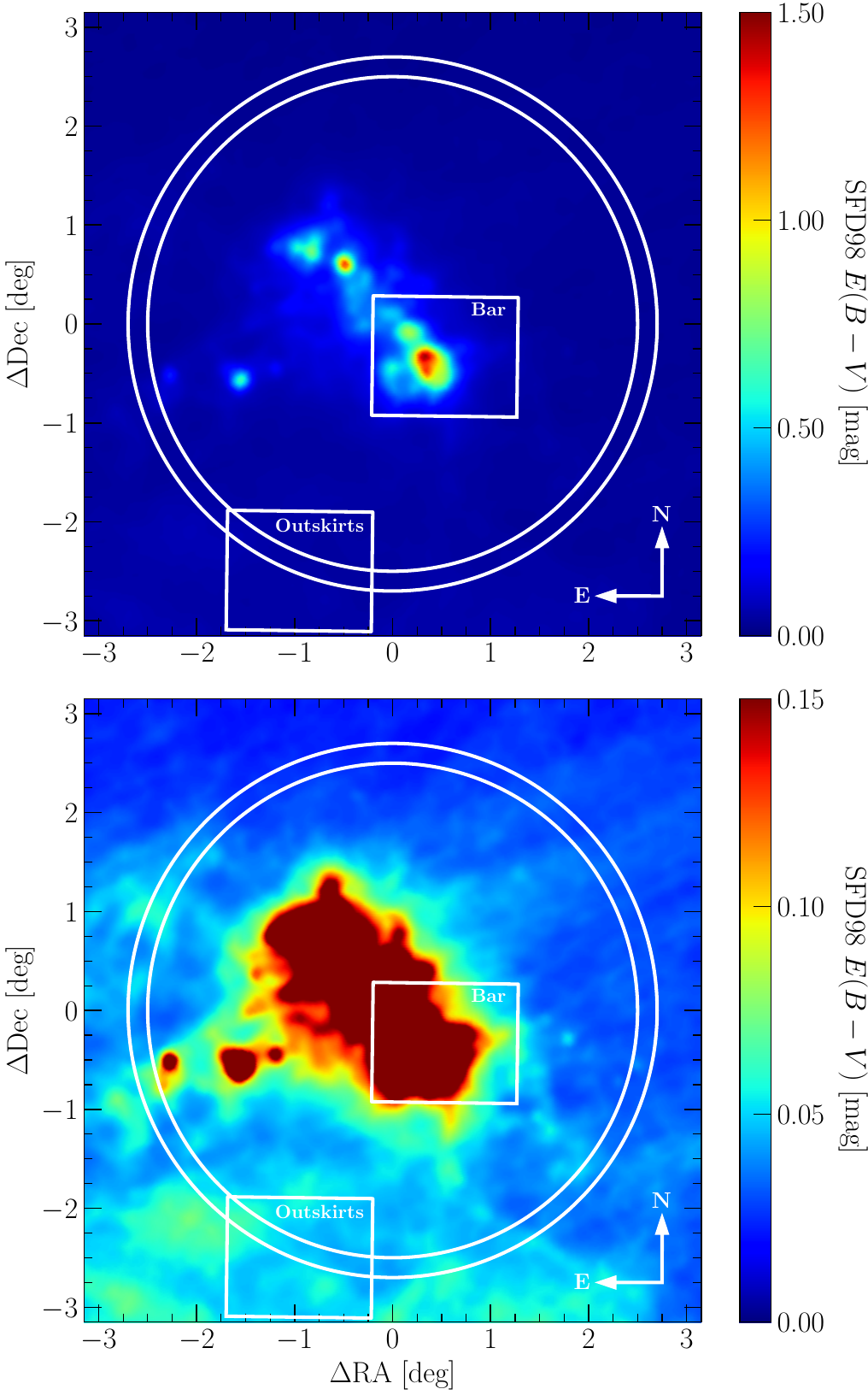}
\caption[]{\cite{Schlegel98} foreground MW and residual intrinsic SMC reddening maps in the direction of the SMC. \emph{Top panel}: The white rectangles denote the positions
of the two tiles studied, whereas the white circles represent the annulus used to determine the median reddening in
the vicinity of the SMC (see text). \emph{Bottom panel}: As the top panel, but plotted using a narrower range of $E(B-V)$ values
to highlight the low-level variations in the foreground and intrinsic reddening across the SMC.}
\label{fig:schlegel_maps}
\end{figure}

Fig.~\ref{fig:schlegel_maps} shows the SFD98 foreground MW and residual intrinsic SMC reddening in the direction of the SMC. 
From the top panel of Fig.~\ref{fig:schlegel_maps} it is clear that towards the Bar and Wing regions
of the SMC the prescription adopted by SFD98 to convert the dust measurements into a foreground reddening
becomes unreliable due to contamination from SMC sources (or unresolved temperature structure; see Appendix~C of SFD98) leading to foreground
$E(B-V)$ values of almost 1.5\,mag in places. Whilst the top panel of Fig.~\ref{fig:schlegel_maps} appears to suggest that in the outer environs of
the SMC the MW foreground reddening is essentially uniform, the bottom panel of
Fig.~\ref{fig:schlegel_maps} shows significant amounts of low-level variation in the SFD98 $E(B-V)$ values.
Regions in the north and west of the panel have systematically lower
reddening values than regions in the south and east. Whether these variations are indeed genuine variations in the foreground
MW reddening or are instead variations in the intrinsic SMC reddening is unclear.

In an attempt to account for the foreground MW reddening in a reasonable and systematic way, we estimate the 
median foreground reddening in the direction of the SMC by
defining an annulus of width 0.2$^{\circ}$ at a radius of 2.5$^{\circ}$ from our adopted SMC centre (see Fig.~\ref{fig:schlegel_maps}).
Using the VMC PSF catalogues, we extract all sources which
lie within this annulus ($\simeq$\,1.9\,$\times 10^{6}$ objects) and use \textsc{sfdmap} to determine the corresponding SFD98 reddening values,
from which we determine a median reddening and standard deviation of $E(B-V)=0.034$\,mag and 0.011\,mag, respectively. Finally, we then
apply this median value to all objects in our photometric catalogues and correspondingly de-redden the individual fluxes
using the coefficients noted in Eqn.~\ref{eqn:ext_coeffs}.

\section{Fitting the SEDs of galaxies}
\label{fitting_seds_of_galaxies}

Whilst there are many available sophisticated codes for modelling
and interpreting galaxy SEDs (see \url{http://www.sedfitting.org/} for a compilation of these), we adopt the
\textsc{lephare} code\footnote{\url{http://cesam.lam.fr/lephare/lephare}} (developed by S.~Arnouts and O.~Ilbert).
Our choice of SED-fitting code is primarily motivated by our goal to produce
reddening maps of the MCs, but is also influenced by several additional factors. For instance, in addition
to the MC reddening, it would be beneficial to have statistics related to the background galaxies, such as
redshift and galaxy type, which could then be compared to other studies.
Furthermore, given that our photometric catalogues still include some
stellar contaminants which we would like to remove, the SED-fitting code needs to be flexible enough
to allow us to fit not only galaxy templates, but also different sets of templates to the SEDs and determine
which provides a better fit to the data. \textsc{lephare} is a $\chi^{2}$ minimisation template-fitting routine
which allows us to fit distinct sets of templates (including QSOs and stars) and thus easily determine the best-fitting model for a given
SED\footnote{In practice it can be hard to reliably differentiate between galaxies and QSOs based only on the $\chi^{2}$ value of
the fit as this can classify a significant number of normal galaxies as QSOs (O.~Ilbert priv. comm.). Thus whilst in some studies only galaxy
and stellar templates are considered (see e.g. \citealp{Ilbert06,Ilbert09}), we chose to include QSO templates as these may provide
a better fit to the SED than those in our galaxy template sets. We therefore note that the numbers of classified galaxies and QSOs listed in
the text should be treated with caution.}.
Importantly, \textsc{lephare} permits inclusion of the intrinsic reddening of background galaxies
[expressed in terms of $E(B-V)$] to vary as a free parameter and explicitly accounts for intergalactic reddening following
the prescription of \cite{Madau95} in which a detailed model
of the H\,{\footnotesize{I}} opacity of the Universe is constructed by analysing the effects of
various absorption mechanisms on the broadband colours of cosmologically distant galaxies.

\subsection{SED template libraries}
\label{sed_template_libraries}

To investigate to what extent the derived reddening values are template-dependent, we adopt two distinct sets of
galaxy templates; one theoretical and the other empirical.
Our base set of theoretical spectra (see \citealp{Ilbert09}) comprises the three elliptical and six spiral
(S0, Sa, Sb, Sc, Sd, Sdm) SEDs generated by \cite{Polletta07} using the \textsc{grasil} code
\citep{Silva98}. \cite{Ilbert09} found that these templates were unable to fully reproduce
the bluest colours of the $z$COSMOS sample of galaxies \citep{Lilly07} and so included
an additional 12 starburst templates generated from the \cite{Bruzual03} models ranging in time of the most recent starburst
from 0.03--3\,Gyr. Finally, \cite{Ilbert09} linearly interpolated
some of the \cite{Polletta07} templates, so our final set of theoretical templates consists of 31 galaxy
SEDs; seven elliptical, 12 spiral and 12 starburst. Our base set of empirical templates
comprises six spectra including the four observed spectral types
(Ell, Sbc, Scd, Irr) from \cite*{Coleman80} in addition to two starburst galaxies from \cite{Kinney96}. These templates
were first extrapolated beyond 1\,$\mu$m using the \textsc{gissel} synthetic models of \cite{Bruzual03} and
then optimised using the rest-frame fluxes of galaxies with spectroscopic redshift determinations in
the VIMOS VLT Deep Survey (VVDS; see \citealp{Ilbert06} for details). Finally, to refine the sampling in
colour-$z$ space, pairs of templates were linearly interpolated to create a final set of 62 SEDs.
The sets of 31 theoretical and 62 empirical templates are referred to as COSMOS and AVEROI\_NEW
in \textsc{lephare} and shall henceforth be referred to as such.

In addition to our galaxy templates, we also adopt two further sets of templates corresponding to QSOs and stars.
The QSO templates consist of 10 SEDs taken from the study of \cite{Polletta07} which includes spectra of moderately
luminous active galactic nuclei (AGN) representing Seyfert 1.8 and Seyfert 2 galaxies, optically selected QSOs with
different values of IR/optical flux ratios, type 2 QSOs as well as composite AGN+starburst galaxies (see \citealp{Polletta07}
for details). The stellar templates comprise a total of 235 spectra including the 131 spectra encompassing all normal spectral
types and luminosity classes at solar abundance as well as metal-rich and metal-poor F--K dwarfs and G--K giants
from \cite{Pickles98}, 100 theoretical spectra of late-type dwarfs of different effective temperatures and
surface gravities from \cite{Chabrier00}, and four white dwarf spectra from \cite*{Bohlin95}. Although the stellar templates do
not include inherently dusty objects such as Carbon and asymptotic giant branch stars, our criteria to select potential galaxies
(see Section~\ref{identification_of_galaxies}) should ensure that these are not an issue.

To compute the fluxes for our templates in the combined DECam $ugriz$ and VIRCAM
$YJK_{\rm{s}}$ bandpasses, all templates were convolved with the corresponding system
responses which are available at the NOAO and ESO webpages, respectively\footnote{The DECam responses are available at
\url{http://www.ctio.noao.edu/noao/content/DECam-filter-information}, whereas the VIRCAM responses are
available at \url{http://www.eso.org/sci/facilities/paranal/instruments/vircam/inst.html}. Note that as the absolute
calibration/normalisation of the DES DR1 bandpasses is not yet available on the DECam webpages, we have
instead adopted the older system responses which are still provided for legacy use.}.
Emission line fluxes can have a significant effect on the magnitudes and colours of galaxies as a function
of redshift, especially if there is significant ongoing star formation (see e.g. \citealp{Jouvel09}). \textsc{lephare}
allows such fluxes to be included in the theoretical magnitudes derived from the galaxy templates and so in order
to better reproduce the colours of such galaxies we choose to include these (see \citealp{Ilbert09} for details
regarding the implementation).

\subsection{Redshift prior}
\label{redshift_prior}

Although the combination of optical and near-IR bands significantly reduces the intrinsic scatter of derived
photometric redshifts as well as the rate of catastrophic failures (see e.g. \citealp{LeFevre13,Wright18}),
the colours of the adopted templates can be degenerate with redshift. To minimise spurious likelihood peaks
in the redshift distribution, we include a Bayesian redshift prior which describes the redshift probability
distribution for galaxies as a function of spectral type (see e.g. \citealp{Benitez00,Ilbert06}) and which
has been calibrated using the robust VVDS spectroscopic sample \citep{LeFevre05}.

\subsection{Extinction law}
\label{extinction_law}

\cite{Ilbert09} used the $z$COSMOS galaxy sample \citep{Lilly07} to test different extinction laws to determine the
best extinction law to adopt for a given spectral type. Their analysis demonstrated that the extinction law as measured
for the SMC by \cite{Prevot84} is well suited for ellipticals, spirals and a subset of starburst galaxies, whereas for
bluer starburst galaxies the \cite{Calzetti00} extinction law as determined from observations of starburst
galaxies is more appropriate. The extinction law for the two starburst galaxies included in the original six spectra from which the
AVEROI\_NEW template set was created is best described by the \cite{Prevot84} formalism and thus we adopt this law
for the AVEROI\_NEW templates. For the COSMOS templates, however, we follow the 
prescription described by \cite{Ilbert09} regarding which extinction law to use for a particular
galaxy template. For our QSO templates we also adopt the \cite{Prevot84} extinction law as this has been found
to reproduce the dust reddening of most QSOs at both high and low redshifts (see e.g. \citealp{Gallerani10}).
To limit degeneracies in the best-fitting solutions (see e.g. \citealp{Arnouts07}), we allow the reddening to vary for the galaxy and QSO templates
from $E(B-V)=0$ to 0.5\,mag in steps of $\Delta E(B-V)=0.05$\,mag.

It is worth stressing that all adopted galaxy and QSO templates already include 
a certain level of dust extinction, either due to a formulated prescription in the code used to create the SEDs
or due to the fact that they are empirical templates. So, even if highly obscured
objects are present in our sample, a combination of the template including some extinction and the possibility to include additional
extinction should ensure such objects are well fitted.

\subsection{Systematic offsets}
\label{systematic_offsets}

Potential uncertainties in both the absolute calibration of the photometric zero points as well as the
predicted fluxes of the template SEDs, resulting from the adopted bandpass response functions
and/or an incomplete template set, can result in zero point offsets between the predicted and observed
colours (see e.g. \citealp{Ilbert06}). The standard approach to account for such uncertainties is
to use a large representative sample of galaxies with spectroscopic redshifts
(i.e. covering the same part of the sky and with photometry using the same instrumental set-up)
and iteratively calculate the required correction factor to ensure that the difference between the predicted
and observed magnitudes at a given redshift is less than a few mmag (see e.g. \citealp{Capak04,Ilbert09}).
Unfortunately, such a representative sample of galaxies
behind the SMC does not exist, and although there are on the order of 200 QSOs (e.g. \citealp{Kozlowski13,Ivanov16})
which we could potentially use, the patchy, non-uniform reddening associated with the SMC makes a
robust determination of zero point offsets difficult to ascertain. Furthermore, given that we would need to
use a template set tailored for QSOs, there is no guarantee that the derived zero point offsets for the
QSOs are necessarily those required by our adopted galaxy template sets. In the near future we expect
large samples of spectroscopic redshifts for galaxies behind the MCs to be obtained with
wide-field spectrographs currently under development (e.g. 4MOST; \citealp{deJong16}).

The recent work on calibrating the VISTA photometric system by \cite{Gonzalez-Fernandez18} shows that
updates to the Cambridge Astronomical Survey Unit (CASU) pipeline can affect the absolute calibration of
the $YJK_{\rm{s}}$ zero points at $\leq 0.02$\,mag level. Although no similar analysis has been performed
for the SMASH data, the absolute photometric calibration of the DES DR1 $griz$ zero points suggests systematic
uncertainties at the $\lesssim$\,0.03\,mag level, although additional sources of systematic uncertainty cannot
be excluded and are currently being investigated \citep{Abbott18}.
Thus in an attempt to account for these zero point offsets, as well as potential uncertainties related
to our choice of template sets, we include
an additional systematic uncertainty of 0.1\,mag in each of our eight bands which is added
in quadrature to the measured uncertainty in each band as calculated from our \textsc{lambdar} photometry
(see Section~\ref{lambdar_photometry}). Note that the amplitude of our offset is motivated by previous studies
in which it has been demonstrated that typical offset amplitudes are $\lesssim$\,0.1\,mag in the optical/near-IR
wavelength regime our data cover (see e.g. \citealp{Brodwin06,Ilbert06,Ilbert09}).

\section{Determining the intrinsic reddening}
\label{determining_intrinsic_reddening_magellanic_clouds}

\begin{table*}
\caption[]{Number of objects classified as a galaxy, QSO or star by \textsc{lephare}.
The values in parentheses denote the corresponding percentage of the total
number of objects in the full sample (see Table~\ref{tab:catalogue_properties}).}
\begin{tabular}{l c c c c c c c}
\hline
   &   &   \multicolumn{3}{c}{COSMOS}   &   \multicolumn{3}{c}{AVEROI\_NEW}\\
Region   &   Sample   &   Galaxies   &   QSOs   &   Stars   &   Galaxies   &   QSOs   &   Stars\\
\hline
\multirow{2}{*}{Outskirts}   &   Aper.   &   12,187 (41.1)   &   16,188 (54.7)   &   1,231 (4.2)   &   7,585 (25.6)   &   20,542 (69.4)   &   1,479 (5.0)\\
   &   PSF   &   15,158 (50.9)   &   13,791 (46.3)   &   839 (2.8)   &   9,904 (33.3)   &   18,838 (63.2)   &   1,046 (3.5)\\
\multirow{2}{*}{Bar}   &   Aper.   &   3,861 (23.0)   &   10,433 (62.1)   &   2,502 (14.9)   &   2,770 (16.5)   &   11,139 (66.3)   &   2,887 (17.2)\\
   &   PSF   &   5,272 (28.9)   &   10,406 (57.0)   &   2,570 (14.1)   &   3,440 (18.9)   &   11,522 (63.1)   &   3,286 (18.0)\\
\hline
\end{tabular}
\label{tab:method1_lephare_class}
\end{table*}

We run \textsc{lephare} on both the standard and PSF-weighted aperture samples of galaxies
for the bar and outskirts regions of the SMC allowing the redshift to vary from $z=0.0$ to 6.0 in steps of $\Delta z=0.02$
(see Table~\ref{tab:method1_lephare_class} for the full classification of all objects in the various samples).
Unlike in typical extragalactic studies, where additional reddening is only implemented for galaxy templates
of type Sc and bluer/later (see e.g. \citealp{Ilbert09}), we are dealing with an extra component of the line-of-sight reddening
(the SMC itself). Thus to ensure that our fits, especially for redder/earlier galaxies, are unbiased, we allow additional
reddening for all galaxy types. In this Section we will first introduce the reddening maps inferred from our preferred
combination of parameters, namely the use of standard apertures, empirical templates and galaxies with low levels of intrinsic
reddening, and then provide a discussion on the systematics introduced when using different combinations of parameters.

\subsection{Reddening maps of the bar and outskirts of the SMC}
\label{refined_sample_galaxies_preferred_combination_parameters}

It is common practice when implementing \textsc{lephare} to only allow additional reddening for galaxy templates corresponding to spectral types
of Scd and bluer/later (see e.g. \citealp{Ilbert06,Arnouts07}). Presumably this is due to the fact that the typically low/negligible
levels of intrinsic reddening inherent to galaxies of spectral types Sbc and redder/earlier are adequately accounted for in the empirical templates (see
e.g. \citealp{Ilbert13}). If we similarly assume that no additional reddening is required for spectral types Sbc and redder/earlier using the AVEROI\_NEW templates,
this would imply that, in our implementation of \textsc{lephare} where we have allowed the reddening to vary in this spectral type range, the best-fitting
$E(B-V)$ values correspond to the total intrinsic reddening of the bar and outskirts regions of the SMC.

\begin{table}
\caption[]{Number of objects which satisfy the three criteria designed to refine the samples. The values in parentheses denote the corresponding percentage of
the total number of objects in the full sample (see Table~\ref{tab:catalogue_properties}).}
\begin{tabular}{l c c c}
\hline
Region   &   Sample   &   COSMOS   &   AVEROI\_NEW\\
\hline
\multirow{2}{*}{Outskirts}   &   Aper.   &   15,031 (50.8)   &   14,537 (49.1)\\
   &   PSF   &   19,812 (66.5)   &   19,124 (64.2)\\
\multirow{2}{*}{Bar}   &   Aper.   &   6,661 (39.7)   &   6,365 (37.9)\\
   &   PSF   &   8,488 (46.5)   &   7,906 (43.3)\\
\hline
\end{tabular}
\label{tab:constraints}
\end{table}

Before we infer the intrinsic reddening maps from the galaxies classified as having spectral types of Sbc and redder/earlier according to the AVEROI\_NEW
templates, and for which the fluxes have been extracted using standard (circular) apertures, there are a few effects that are worth considering which could
potentially affect the resultant reddening, namely the number of bandpasses in the SEDs, blending and incompleteness issues at faint magnitudes and redshift
probability distributions with multiple peaks. Placing constraints on these would provide the most robust sample from which to infer the reddening maps.
Appendix~\ref{refining_sample_galaxies} addresses each of these points in detail and Table~\ref{tab:constraints} lists the number of objects which pass these
three selection criteria (having removed objects classified as stars).

\begin{figure}
\centering
\includegraphics[width=\columnwidth]{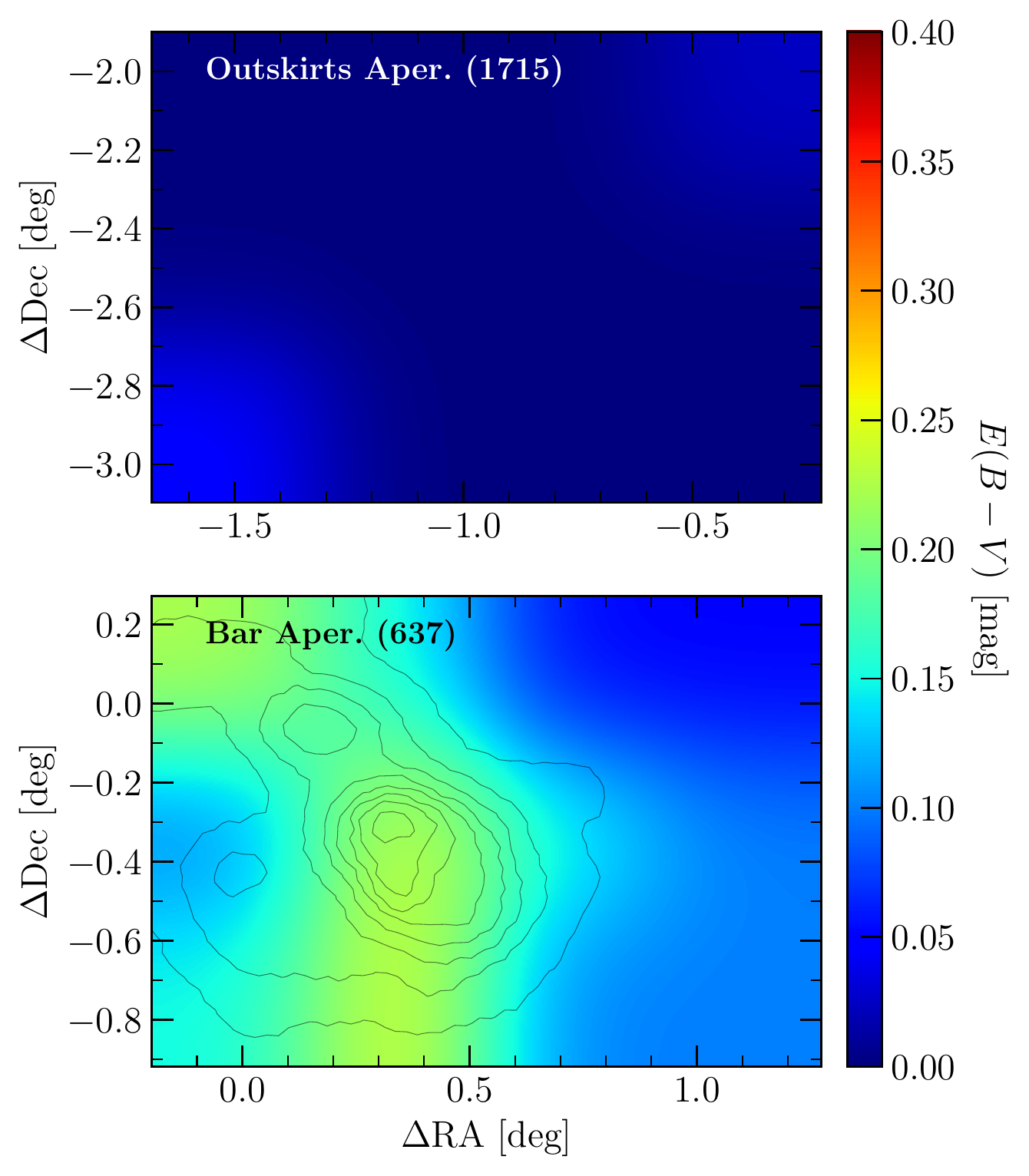}
\caption[]{Gaussian smoothed $20 \times 20$\,arcmin$^{2}$ resolution reddening maps for the standard aperture
samples of the outskirts (top) and bar (bottom) regions of the SMC. These maps have been created using only galaxies with
spectral types of Sbc and redder/earlier according to the AVEROI\_NEW templates and which satisfy the three criteria
designed to refine the sample (see text). The values in parentheses denote the number of galaxies in each sample and the
contours in the bottom panel represent the IRAS 100\,$\mu$m dust emission.}
\label{fig:red_maps_averoin_gal_passive_stan_aper_refined}
\end{figure}

Fig.~\ref{fig:red_maps_averoin_gal_passive_stan_aper_refined} shows $20 \times 20$\,arcmin$^{2}$ resolution reddening maps, smoothed with an
8\,arcmin Gaussian, for the standard aperture samples of the outskirts and bar regions of the SMC using only objects classified by \textsc{lephare} as galaxies
with spectral types of Sbc and redder/earlier according to the AVEROI\_NEW templates and which satisfy the three criteria designed to refine the sample
(see Appendix~\ref{refining_sample_galaxies}). To determine the $E(B-V)$  value within a given bin, we take the median of all best-fitting $E(B-V)$ values.
The number of galaxies in a given bin varies from 87 to 217 for the outskirts region and 16 to 125 for the bar
region (likely as a result of differing limiting magnitudes between the two regions; see Appendix~\ref{blending_incompleteness}), and the median reddening values
across both regions are $E(B-V)_{\mathrm{med}}=0.0$ and 0.13\,mag, respectively. The associated uncertainty on the reddening
values ranges from $\sigma_{E(B-V)_{\mathrm{med}}}=0.07$ to 0.10\,mag and from $\sigma_{E(B-V)_{\mathrm{med}}}=0.10$ to 0.16\,mag for the outskirts and bar regions,
respectively. These values represent the uncertainties in a given bin and are calculated as the standard deviation of the galaxy reddening values within that
bin. To help the reader visualise where the majority of the dust lies in the central regions of the SMC we overlay the IRAS 100\,$\mu$m emission contours in the bottom
panel of Fig.~\ref{fig:red_maps_averoin_gal_passive_stan_aper_refined}. Note that due to the low levels of intrinsic dust emission, in addition to a lack of variation,
across the outskirts region of the SMC, we only plot the emission contours in the bar region.

\subsection{Effects of adopting different templates, apertures and galaxy samples}
\label{effects_adopting_different_galaxy_samples_templates}

In Section~\ref{refined_sample_galaxies_preferred_combination_parameters} we presented intrinsic reddening maps of the SMC based on our preferred combination of
parameters, as well as three criteria designed to refine the sample of galaxies from which the maps were inferred. In this Section we will investigate what effect adopting
different templates, apertures and galaxy samples would have on the inferred maps. Fig.~\ref{fig:red_maps_all} presents a large number of inferred reddening maps which
will form the basis of our discussion regarding these effects. To quantify how each individual effect affects the inferred reddening map, we will adopt the combination
of parameters used to create Fig.~\ref{fig:red_maps_averoin_gal_passive_stan_aper_refined} as our reference.

\subsubsection{Empirical vs. theoretical templates}
\label{choice_galaxy_template}

We first investigate the effect of adopting theoretical galaxy templates as opposed to empirical templates. Fig.~\ref{fig:red_maps_all} clearly shows that the
use of theoretical templates results in increased levels of reddening, regardless of galaxy sample, aperture, etc. We determine an increase in the median inferred
reddening of $E(B-V)_{\mathrm{med}}=0.14$ and 0.20\,mag for the outskirts and bar regions, respectively. These values are larger than the associated uncertainties
($\sigma_{E(B-V)_{\mathrm{med}}}\simeq$\,0.1
and 0.14\,mag, respectively) and thus we conclude that the use of the COSMOS theoretical templates results in inferred reddening
maps with statistically significant enhanced levels of reddening.

\subsubsection{Standard vs. PSF-weighted apertures}
\label{standard_psf_apertures}

As discussed in Section~\ref{lambdar_photometry}, the use of PSF-weighted apertures can affect the resultant SED compared to if fluxes are extracted using
standard apertures. From Fig.~\ref{fig:red_maps_all}, we note a statistically insignificant (given
the associated uncertainties) increase in the median reddening of $E(B-V)_{\mathrm{med}}=0.03$ and 0.05\,mag for the outskirts and bar regions, respectively, as a result of adopting
PSF-weighted apertures.

\subsubsection{Galaxy samples}
\label{galaxy_samples}

In the preceding Sections we have only discussed the use of galaxies with low levels of intrinsic reddening to create intrinsic reddening maps of the two regions
of the SMC. Such galaxies constitute only a small fraction of the full sample of background objects (both galaxies and QSOs) and so here we investigate what effects
increasing our sample size (including the addition of dusty irregular and/or starburst galaxies) has on the inferred reddening maps. From Fig.~\ref{fig:red_maps_all} we
can see that the inclusion of all potential background objects markedly alters the inferred reddening maps, such that both regions show similar levels of intrinsic reddening.
Although not shown in Fig.~\ref{fig:red_maps_all}, similar results are found when creating maps using only galaxies and only QSOs.
The median reddening for both regions is $E(B-V)_{\mathrm{med}}=0.15$\,mag, which equates to an increase of $E(B-V)_{\mathrm{med}}=0.15$ and 0.02\,mag for the
outskirts and bar regions,
respectively. Such a finding is in stark contrast to what we infer using only galaxies with low levels of intrinsic reddening as well as previous reddening maps of the SMC
based on its stellar components or dust emission, both of which clearly demonstrate that the bar region shows enhanced levels of reddening compared to the outer environs
(see Section~\ref{discussion}).

Note that for the COSMOS templates we have adopted two distinct extinction laws depending on the spectral type of the template
(see Section~\ref{extinction_law}). Thus one can argue that to infer the intrinsic reddening of the SMC using all background objects, all templates should have been
treated using the same \emph{single} SMC-like extinction law. Given the similarity between the \cite{Prevot84} and \cite{Calzetti00} extinction laws, in conjunction with
the low levels of intrinsic reddening inherent to the SMC and the discretisation of reddening values adopted in our implementation of \textsc{lephare} [$\Delta E(B-V)=0.05$\,mag],
we find that this inconsistency has negligible effect on the inferred reddening maps. Note that as all AVEROI\_NEW templates were treated with the same SMC-like
extinction law, our preferred reddening maps shown in Fig.~\ref{fig:red_maps_averoin_gal_passive_stan_aper_refined} do not suffer from this inconsistency.

\subsubsection{Criteria designed to refine the sample}
\label{criteria_designed_refine_sample}

\begin{landscape}
\begin{figure}
\centering
\includegraphics[width=\linewidth]{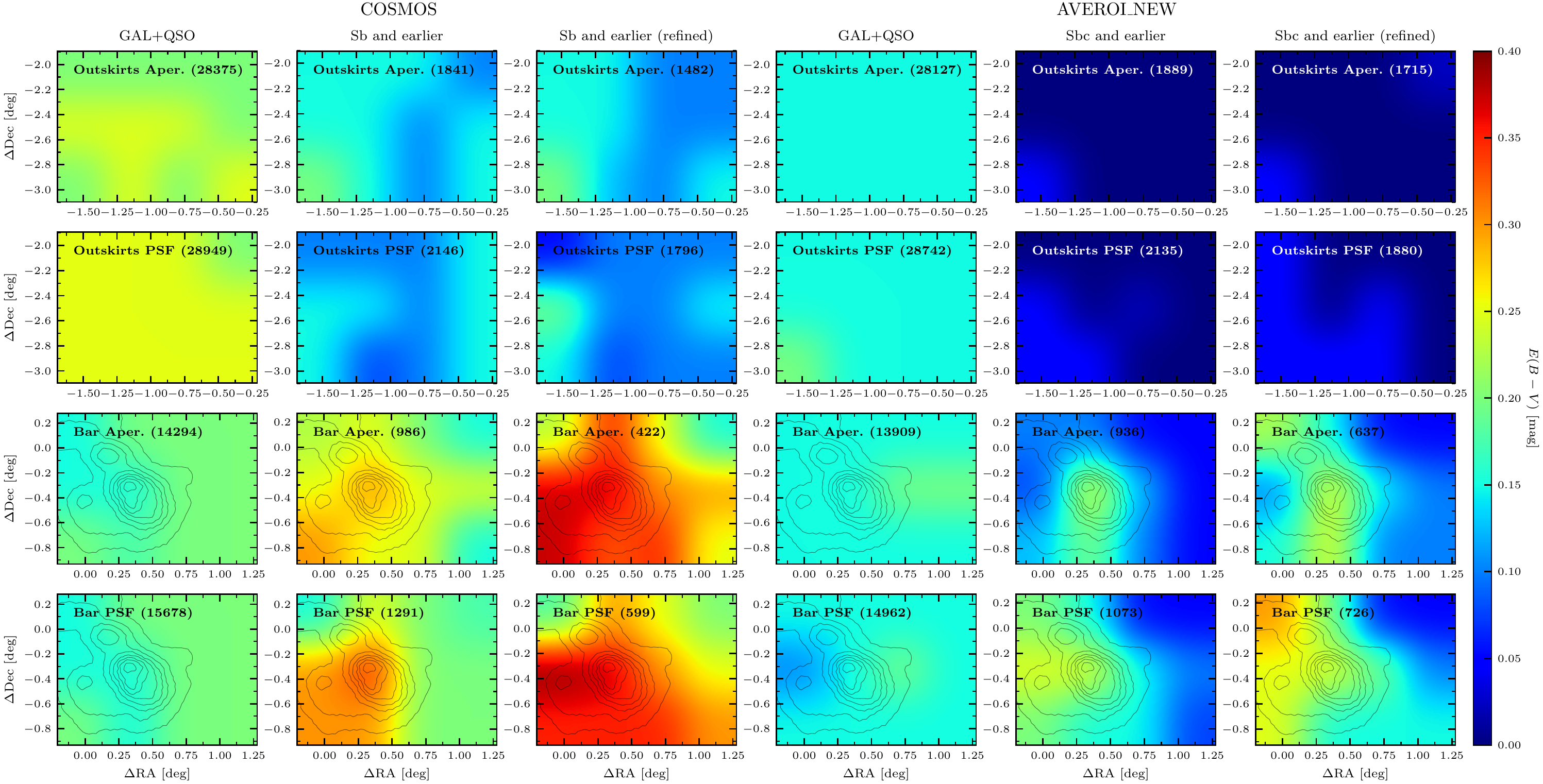}
\caption[]{Gaussian smoothed $20 \times 20$\,arcmin$^{2}$ resolution reddening maps for the outskirts and bar regions of the SMC showing the effects introduced as a
result of adopting i) different galaxy templates, ii) different apertures to extract the fluxes, iii) different samples of galaxies and iv) selection criteria designed to refine
the sample of galaxies. The first three columns are based on the theoretical COSMOS templates, whereas the final three are based on the empirical AVEROI\_NEW
templates. The samples used to infer the reddening maps are as follows: first column (full sample of galaxies and QSOs), second column [galaxies with spectral types
of Sb (Sbc) and redder/earlier according to the COSMOS (AVEROI\_NEW) templates] and third column (same as second column, but also incorporating three criteria
designed to further refine the sample of galaxies; see text). The contours in the bottom two rows represent the IRAS 100\,$\mu$m dust emission.}
\label{fig:red_maps_all}
\end{figure}
\end{landscape}

Finally, we investigate what effect the selection criteria designed to refine the sample of galaxies has on the inferred reddening maps. From Fig.~\ref{fig:red_maps_all}
we note a small change in the physical appearance of the reddening maps for both regions. The median reddening, however, only decreases by
$E(B-V)_{\mathrm{med}}=0.03$\,mag in the bar region
as a result of not including the selection criteria (the median reddening of the outskirts region is unchanged). Note that when using the COSMOS templates, the median reddening
can vary by $E(B-V)_{\mathrm{med}}\simeq$\,0.1\,mag due to the selection criteria alone, although this is still within the associated uncertainties and thus we
conclude that the inclusion of the selection criteria has a statistically insignificant effect on the resultant reddening maps.

\subsection{Galaxy spectral type and photometric redshift distributions}
\label{galaxy_template_photo_redshift_distributions}

\begin{figure}
\centering
\includegraphics[width=\columnwidth]{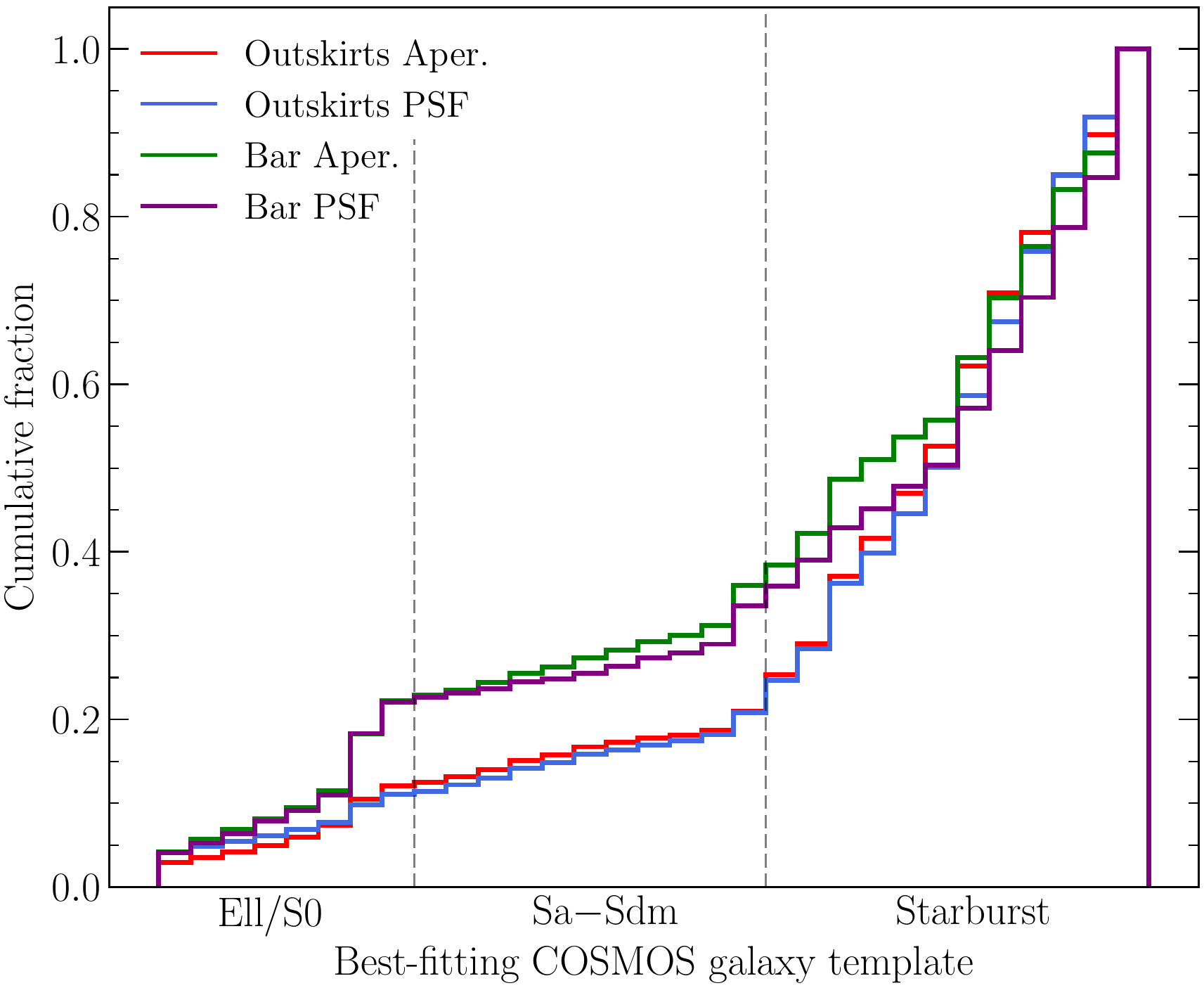}
\caption[]{Normalised cumulative distributions showing the distribution of best-fitting galaxy templates (elliptical/lenticular, spiral and starburst) for 
galaxies (based on the COSMOS templates) in the standard and PSF-weighted aperture samples for both SMC tiles.}
\label{fig:method1_galaxy_dist_cosmos_qso_comb}
\end{figure}

\begin{figure}
\centering
\includegraphics[width=\columnwidth]{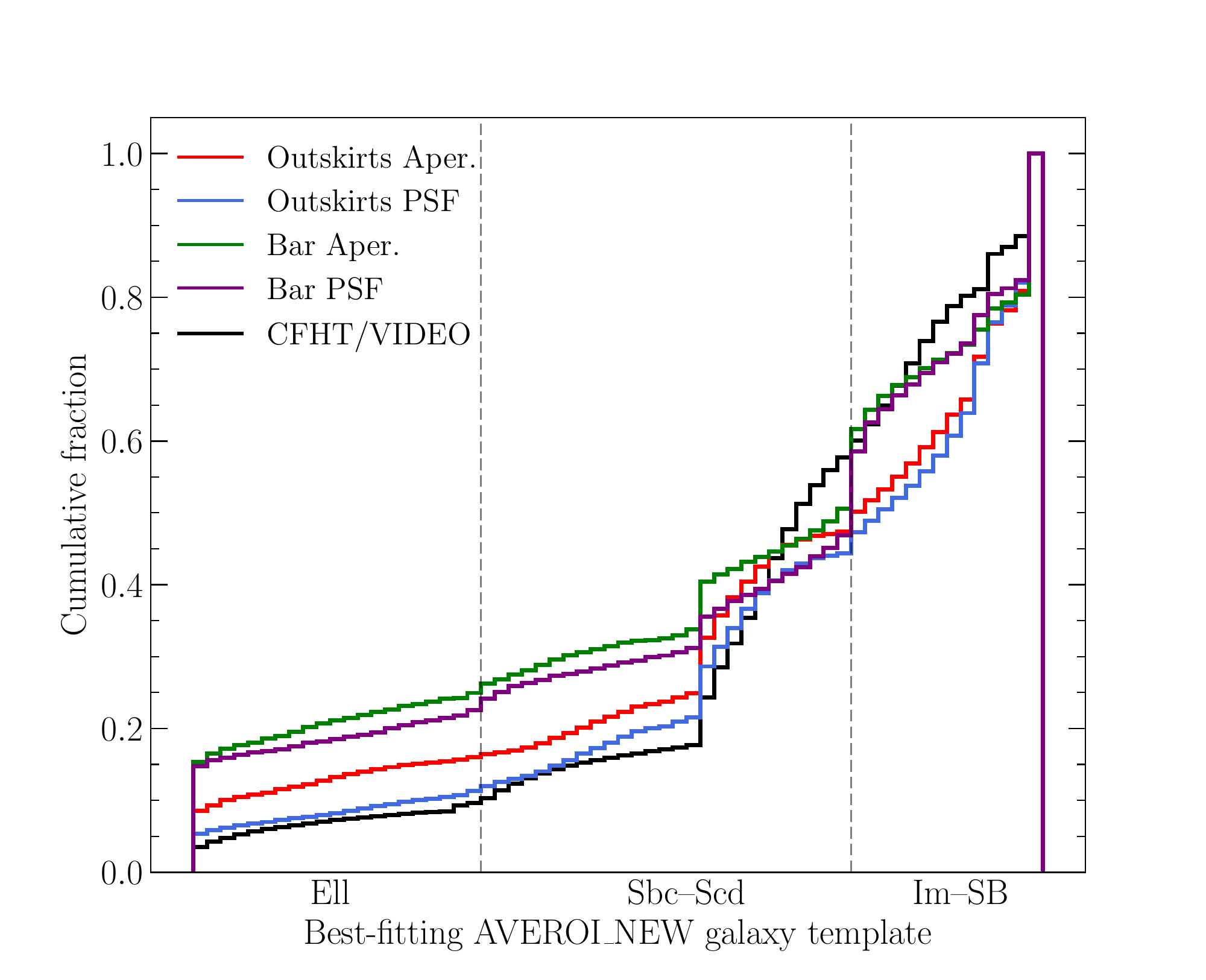}
\caption[]{As Fig.~\ref{fig:method1_galaxy_dist_cosmos_qso_comb}, but based on the AVEROI\_NEW templates and grouped into elliptical, spiral
and irregular/starburst. The distribution for the CFHT/VIDEO sample (see text for details) is also shown.}
\label{fig:method1_galaxy_dist_averoin_qso_comb}
\end{figure}

Although studies of galaxies behind the MCs are not new, this study represents (one of) the first to investigate them in large numbers and thus some discussion of the
galaxy statistics is warranted. Figs.~\ref{fig:method1_galaxy_dist_cosmos_qso_comb} and \ref{fig:method1_galaxy_dist_averoin_qso_comb} show the distribution of best-fitting
galaxy templates based on the COSMOS and AVEROI\_NEW templates, respectively.
These figures highlight that although the distributions for a given region, irrespective of whether the fluxes have been measured using standard or PSF-weighted apertures,
are very similar with comparable numbers of each galaxy type, there are systematic differences in the galaxy type distribution between the bar and outskirts regions. It is likely
that a combination of the difference in the line-of-sight reddening and crowding between the two regions plays a significant role in the resultant differences in
best-fitting galaxy template fractions.

In addition, there is a clear template-dependency on the inferred fraction of galaxy types. Although the fractions of elliptical/lenticular galaxies are similar for both
sets of templates (10--20 per cent depending on the region), there are significant differences in the numbers of both spirals and irregular/starburst galaxies. The AVEROI\_NEW
templates suggest a plausible number of spiral galaxies (30--35 per cent), whereas only $\simeq$\,10 per cent of galaxies are classified as spirals with the COSMOS templates.
Similarly, a staggering 65--80 per cent of galaxies are classified as starburst by the COSMOS templates. Such numbers are likely attributable to how the COSMOS template
set was created, however a comprehensive analysis of these templates is outside the remit of this study.

\begin{figure}
\centering
\includegraphics[width=\columnwidth]{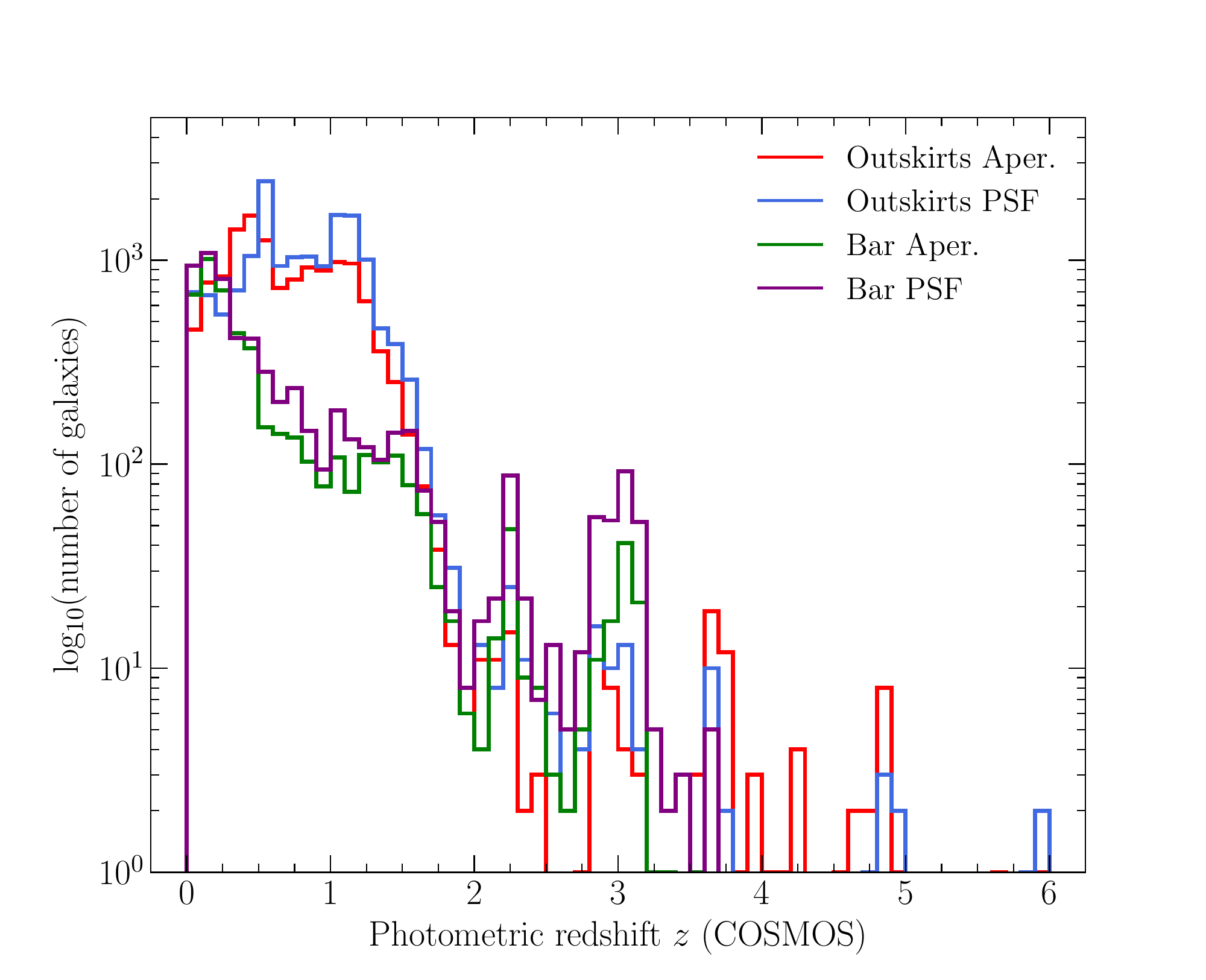}
\caption[]{Photometric redshift distributions of galaxies (based on the COSMOS templates) for both the standard and PSF-weighted
aperture samples for both SMC tiles.}
\label{fig:method1_redshift_dist_cosmos}
\end{figure}

\begin{figure}
\centering
\includegraphics[width=\columnwidth]{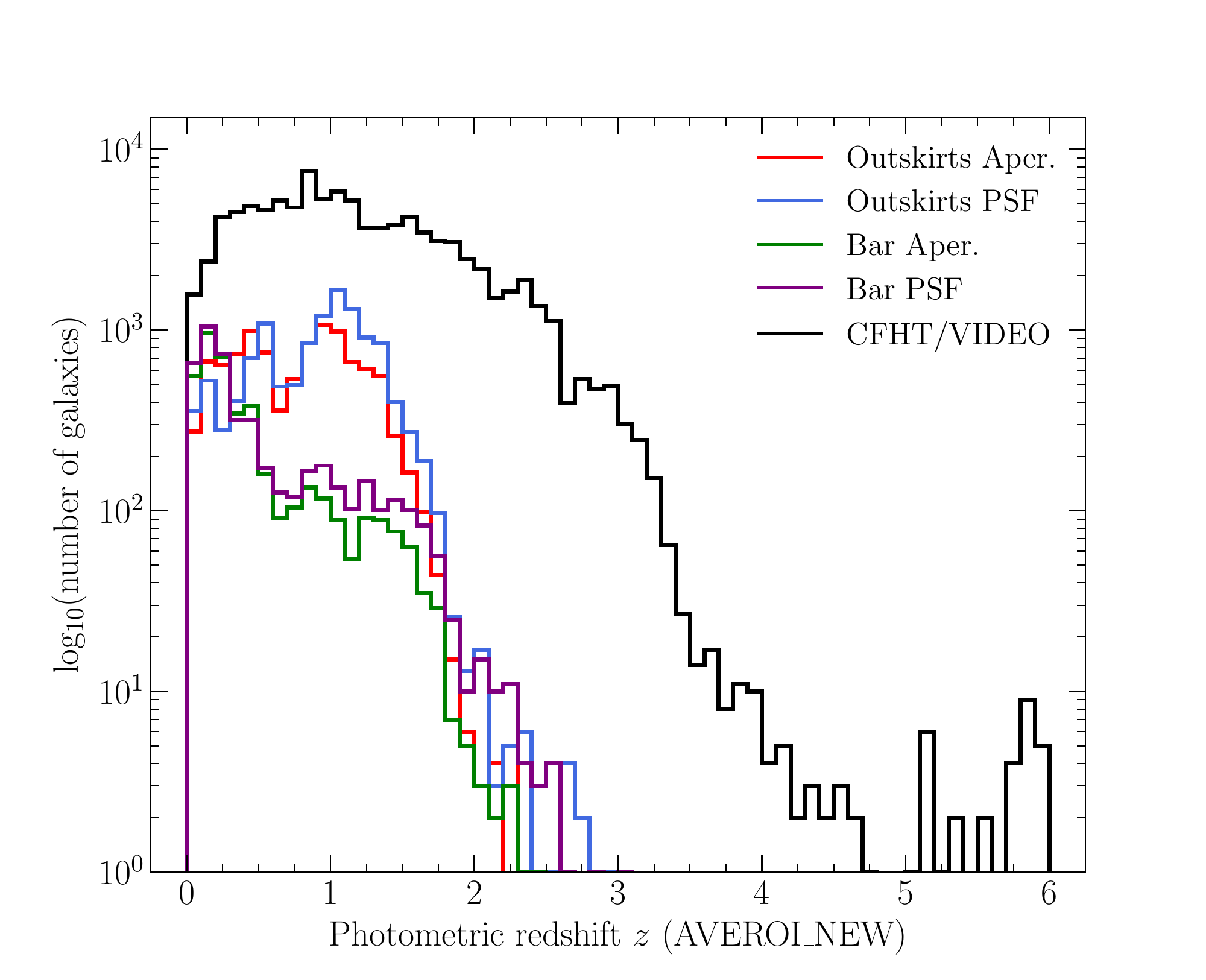}
\caption[]{As Fig.~\ref{fig:method1_redshift_dist_cosmos}, but based on the AVEROI\_NEW templates and also showing the CFHT/VIDEO distribution.}
\label{fig:method1_redshift_dist_averoin}
\end{figure}

Figs.~\ref{fig:method1_redshift_dist_cosmos} and \ref{fig:method1_redshift_dist_averoin} show the photometric redshift distributions of galaxies based on the
COSMOS and AVEROI\_NEW templates, respectively. Irrespective of the adopted template set, the peak in the redshift distribution is essentially the same for
a given sample. Using the AVEROI\_NEW templates, at redshifts higher than $z$\,$\simeq$\,0.2 and 1 (for the bar and outskirts regions, respectively), we
see a gradual decrease in the number of galaxies as a function of redshift, with no galaxies at redshifts beyond $z=3$. In the case of
the COSMOS templates we see a similar redshift distribution as for the AVEROI\_NEW templates up to a redshift $z=2$, however there are also a non-negligible number of
galaxies with higher redshifts in all samples (see e.g. the bump at $z\sim3$ in Fig.~\ref{fig:method1_redshift_dist_cosmos}). Given that we do not see such effects
when using empirical templates, we can only surmise that this results from the use of theoretical templates (see also \citealp{Onodera12} who showed potential
issues with using theoretical templates to determine photometric redshifts). The clear difference in the peak of the redshift distribution between the outskirts and
bar regions, irrespective of adopted templates, suggests that the combination of crowding and incompleteness (see Appendix~\ref{blending_incompleteness}) has
a marked effect on the latter resulting in a peak at much lower redshifts.

\subsection{Comparison with literature results}
\label{comparison_literature_results}

To place our results in context, in Figs.~\ref{fig:method1_galaxy_dist_averoin_qso_comb} and \ref{fig:method1_redshift_dist_averoin} we also plot the
best-fitting galaxy template and photometric redshift distributions, respectively, resulting from the analysis of the combined CFHTLS-DF1 optical $ugriz$ data
and the VISTA Deep Extragalactic Observations (VIDEO) $ZYJHK_{\mathrm{s}}$ data performed by \cite{Jarvis13} and provided by M.~Jarvis
(priv. comm.)\footnote{These data differ somewhat from those published by \cite{Jarvis13} and so differences may be observed between the photometric redshift
distribution shown here and that shown in their fig.~13.}. Given that the wavelength coverage is almost identical between the CFHT/VIDEO sample and ours, as well
as the fact that \cite{Jarvis13} use the \textsc{lephare} code in conjunction with the same galaxy (AVEROI\_NEW)/QSO templates as
used here makes this a favourable data set to compare against. The fraction of ellipticals between the standard and PSF-weighted aperture samples of the outskirts region
and the CFHT/VIDEO sample agree to $\sim$\,2--6\,per cent, whereas this decreases to only $\sim$\,15--17\,per cent for spirals and $\sim$\,10--13\,per cent for irregular and
starburst galaxies. In contrast, for the bar region samples the fractions agree to within $\sim$\,12--15, 14--23 and 7--11\,per cent for elliptical, spiral and irregular/starburst
galaxies, respectively. Although the significantly higher number of high-$z$ galaxies in the CFHT/VIDEO sample (see Fig.~\ref{fig:method1_redshift_dist_averoin}) clearly
demonstrates the deeper photometry in the \cite{Jarvis13} sample, it is reassuring to see that the redshift distributions for the outskirts region peak at a similar redshift.

\begin{table*}
\caption[]{Comparison of the spectroscopically measured and calculated photometric redshifts for QSOs in each sample.}
\begin{tabular}{l c c c c c c}
\hline
   &   &   \multicolumn{2}{c}{COSMOS/QSO photo-$z$}   &   \multicolumn{2}{c}{AVEROI\_NEW/QSO photo-$z$}   &\\
Name   &   Spectro-$z$   &   Aper.   &   PSF   &   Aper.   &   PSF   &   Reference\\
\hline
XMMU J003917.7$-$730330   &   0.345   &   $0.480\pm0.014$   &   $0.500\pm0.014$   &   $0.321\pm0.015$   &   $0.500\pm0.014$   &   1\\
MQS J004143.75$-$731017.1   &   0.217   &   $0.192^{+0.021}_{-0.031}$   &   $0.203\pm0.016$   &   $0.239^{+0.014}_{-0.015}$   &   $0.175^{+0.090}_{-0.023}$   &   2$^{a}$\\
MQS J004145.04$-$725435.9   &   0.267   &   $0.517^{+0.022}_{-0.027}$   &   $0.339^{+0.092}_{-0.025}$   &   $0.517^{+0.022}_{-0.027}$   &   $0.339^{+0.092}_{-0.025}$   &   2\\
OGLE SMC125.2 55437   &   0.785   &   $0.408^{+0.230}_{-0.155}$   &   $0.350^{+0.181}_{-0.129}$   &   $0.441^{+0.199}_{-0.165}$   &   $0.382\pm0.126$   &   3$^{b}$\\
MQS J004736.12$-$724538.2   &   0.572   &   $0.618^{+0.015}_{-0.013}$   &   $0.620\pm0.014$   &   $0.618^{+0.015}_{-0.013}$   &   $0.620\pm0.014$   &   2\\
OGLE SMC100.6 59954   &   2.086   &   $0.179^{+0.572}_{-0.103}$   &   $0.759\pm0.014$   &   $0.750^{+0.027}_{-0.037}$   &   $0.760\pm0.014$   &   3$^{b}$\\
OGLE SMC100.4 26477   &   0.288   &   $0.253^{+0.083}_{-0.028}$   &   $0.161^{+0.017}_{-0.016}$   &   $0.273^{+0.020}_{-0.026}$   &   $0.162\pm0.015$   &   3\\
OGLE SMC105.7 34076   &   0.505   &   $0.480\pm0.014$   &   $0.100\pm0.014$   &   $0.500\pm0.014$   &   --   &   3$^{a}$\\
\hline
\end{tabular}
\vspace{1pt}
\begin{flushleft}
References: (1) \protect\cite{Maitra19}, (2) \protect\cite{Kozlowski13}, (3) \protect\cite{Kozlowski11}.\\
$^{a}$Best-fitting template corresponds to a galaxy template.\\
$^{b}$Listed as plausible QSO due to low-quality spectra.
\end{flushleft}
\label{tab:spec_phot_z}
\end{table*}

As an additional test of our \textsc{lephare} outputs, in Table~\ref{tab:spec_phot_z} we compare the calculated photometric redshifts to spectroscopic redshifts of QSOs in our
two SMC tiles. We combine the spectroscopically confirmed QSOs behind the SMC from \cite{Dobrzycki03a,Dobrzycki03b}, \cite{Geha03}, \cite{Veron-Cetty10},
\cite*{Kozlowski11}, \cite{Kozlowski13}, \cite{Ivanov16}, and \cite{Maitra19}, and identify a total of 33 unique matches within the VMC PSF catalogues of the two SMC tiles adopting
a 1\,arcsec matching radius. Of these all but one lie in the bar region, of which only eight pass our selection criteria.
The remaining QSO in the outskirts region does not pass these selection criteria. Table~\ref{tab:spec_phot_z} shows that there is a good correlation between the
spectroscopic and photometric redshifts, although the uncertainties on the latter do appear to be significantly underestimated.
Of the eight QSOs listed in Table~\ref{tab:spec_phot_z}, for only one all four
photometric redshift determinations are within 3$\sigma$ of the spectroscopic redshift, however for six at least one photometric redshift is consistent
at the same level. Given the small-number statistics, we defer drawing any significant conclusions until the full analysis of the SMC is presented
in a forthcoming paper.


\section{Discussion and comparisons}
\label{discussion}

The foremost conclusion, based on our knowledge of the reddening distribution of the SMC from previous studies, is that one is unable
to produce reliable reddening maps based on all objects classified as galaxies and/or QSOs (see Section~\ref{galaxy_samples}).
It is unclear exactly what the reason for this is. However, the seemingly large number of starburst galaxies present in our samples could play a significant role
(see Figs.~\ref{fig:method1_galaxy_dist_cosmos_qso_comb} and \ref{fig:method1_galaxy_dist_averoin_qso_comb}). Compared to elliptical/lenticular
and early type spirals, starburst galaxies exhibit significantly higher levels of intrinsic reddening. It could be that the additional reddening provided by the SMC
itself acts to change the shape of the spectrum such that it is better fit by a starburst galaxy template. Such starburst templates do not necessarily require additional
reddening to provide a good fit to the data, and thus could act to bias any determination of the reddening of the foreground SMC.

Conversely, if we remove galaxies with high levels of intrinsic reddening (e.g. late-type spirals, irregulars and starburst) then the resultant reddening maps
appear to be more consistent with what we know regarding the reddening distribution of the SMC; namely that, other than the Wing, the Bar exhibits higher
levels of reddening than regions farther from the centre (see e.g. \citealp{Dobashi09,Haschke11,Rubele18}). Note, however that we still see regions within
both regions (irrespective of the adopted templates) which exhibit high levels of intrinsic reddening.  A comparison to previously published reddening maps
allows us to better understand if these are also present in other maps or whether these features potentially indicate enhanced levels of dust and/or could be
attributed to problems associated with our methodology.

In the following we compare our preferred reddening maps (see Section~\ref{refined_sample_galaxies_preferred_combination_parameters}) to reddening maps
from the literature. The publicly available reddening data from which we can construct the comparison
reddening maps are based on various stellar components of the SMC. A significant effect in comparing maps based on background galaxies and stars in the SMC
is the depth effect. Whereas the galaxies trace the full line-of-sight dust column through the SMC, the stars only trace the reddening to some mean distance in the SMC.
If the SMC has a long tidal tail of debris (including dust) that lies primarily along the line of sight (see e.g. \citealp{Jacyszyn-Dobrzeniecka16,Ripepi17}), then the galaxies
will naturally show enhanced level of reddening, compared to the stellar reddening values which sample less of the total column depth. Assuming that the stellar reddening
values represent the ``mean'' reddening along a given line of sight, one could naively argue that these values should be multiplied by a factor of $\simeq$\,2
to account for the ``remaining'' volume of the SMC to be traced. Given the asymmetric depth of the SMC as a function of position (see e.g. \citealp{Subramanian12,Nidever13}),
such an assumption is not necessarily warranted and so we do not apply such a correction but note it for the sake of our comparisons. Another feature to note regarding the
stellar reddening values is that they tend to provide line-of-sight measurements which also include the foreground MW component (something which we have explicitly removed
as we are interested in determining the total intrinsic reddening within the two SMC regions). We therefore remove the foreground MW component from the literature
maps by simply subtracting $E(B-V)=0.034$\,mag (see Section~\ref{foreground_reddening}) from each line-of-sight reddening measurement.

\subsection{SMC reddening maps from the literature}
\label{literature_reddening_maps}

In this Section we briefly introduce the SMC reddening maps from the literature which we include in our comparison to the reddening maps presented here
based on background galaxies (see Fig.~\ref{fig:red_maps_averoin_gal_passive_stan_aper_refined}). The literature maps are not only based on different stellar
components of the SMC, but have also been derived using different wavelength regimes covering the optical and near-IR. To facilitate a comparison, we have
re-sampled the literature maps to the same $20 \times 20$\,arcmin$^{2}$ resolution as adopted for our reddening maps by simply taking the median reddening value
of all stars/cells which fall within a given bin.

\subsubsection{\protect\cite{Zaritsky02}}
\label{zaritsky}

\cite{Zaritsky02} used multi-band optical data from the Magellanic Clouds Photometric Survey (MCPS; \citealp*{Zaritsky97}) to construct 18\,deg$^{2}$ $V$-band extinction maps
of the SMC based on hot ($12,000 < T_{\rm{eff}} < 45,000$\,K) and cool ($5500 < T_{\rm{eff}} < 6500$\,K)
stars\footnote{\url{http://djuma.as.arizona.edu/~dennis/mcsurvey/Data_Products.html}.}. We chose to compare our reddening maps based on background galaxies
to the cool star extinction maps, as although the hot (young) stars will provide a better probe of the dust associated with recent/on-going star formation, any pervasive
dust signatures may be affected by small filling factors due to the biased sampling of such stars. In contrast, the cool stars will provide a more uniform probe of the
reddening across the SMC. We transform $A_V$ into $E(B-V)$ following the standard $A_V = 3.1\times E(B-V)$ \citep{Cardelli89}. The MCPS footprint only covers a very
small fraction of the outskirts region and thus we limit our comparison to the bar region only.

\subsubsection{\protect\cite{Haschke11}}
\label{haschke}

\cite{Haschke11} used $V$- and $I$-band data from the third phase of the Optical Gravitational Lensing Experiment (OGLE III; \citealp{Udalski03}) to create 14\,deg$^{2}$
$E(V-I)$ reddening maps of the SMC based on red clump and RR Lyrae stars, however only the former is publicly
available\footnote{\url{http://dc.zah.uni-heidelberg.de/mcextinct/q/cone/form}}. To transform the $E(V-I)$ values determined by \cite{Haschke11} into $E(B-V)$, we
adopt the prescription of \citet*[$E(V-I)=1.38 \times E(B-V)$]{Tammann03}. Note that the OGLE-III footprint does not include the outskirts region.
Upon subtracting the foreground MW component from the \cite{Haschke11} values, the number of remaining cells with positive reddening values is significantly reduced.
This not only implies that essentially zero intrinsic reddening is associated with the bar region, although in the most south-westerly bin of the bar region there were no cells with
which to calculate a median reddening. We therefore adopted the median reddening from the remaining 11 bins in this region. We note that other studies have also
used OGLE-III data to study the reddening properties of the SMC resulting in very similar reddening maps (see e.g. \citealp{Subramanian12,Nayak18}).

\subsubsection{\protect\cite{Rubele18}}
\label{rubele}

\cite{Rubele18} used near-IR VMC data to determine the star formation history of the central 23\,deg$^{2}$ of the SMC with a spatial resolution of approximately $20 \times 20$\,arcmin.
The derivation of the star formation history consists of determining the linear combination of partial models that best fit the observed $K_{\rm{s}}$, $Y-K_{\rm{s}}$ and
$K_{\rm{s}}$, $J-K_{\rm{s}}$ colour-magnitude Hess diagrams. These partial models include the effects of extinction, and so by finding the best-fitting combination of models,
\cite{Rubele18} determine a representative ``mean'' $V$-band extinction for each subregion. We transform $A_V$ into $E(B-V)$ in the same way as for the
\cite{Zaritsky02} reddening map (see Section~\ref{zaritsky}).

\subsubsection{\protect\cite{Muraveva18}}
\label{muraveva}

\cite{Muraveva18} used a combination of near-IR time-series photometry from the VMC and optical light curves from OGLE IV to study the structure of the SMC as traced by RR Lyrae
stars. As part of this analysis, they also determined optical $E(V-I)$ reddening values by comparing the observed colours of these stars to their intrinsic colour (based on an empirical
relation connecting the intrinsic colour of the star to its $V$-band amplitude and period; see also \citealp{Haschke11}). We transform $E(V-I)$ to $E(B-V)$ using the same
formalism as for the \cite{Haschke11} reddening map (see Section~\ref{haschke}). Importantly, this sample also covers the outskirts region,
which thus provides a means to test our galaxy reddening maps in regions much less affected by crowding and with lower levels of reddening. There are no RR Lyrae from the
\cite{Muraveva18} sample which occupy the most south-easterly bin of the outskirts region and thus for this bin we adopted the median value from the other 11 bins in
this region.

\subsubsection{Tatton et al. (in preparation)}
\label{tatton}

Tatton et al. (in preparation) used near-IR VMC data to construct an $E(Y-K_{\rm{s}}$) reddening map covering  $\simeq$\,45\,deg$^{2}$ of the SMC using red clump stars adopting
a very similar method to those of \cite{Haschke11} and \cite{Subramanian12}\footnote{This work is currently in preparation. The thesis containing it can be accessed at
\url{http://eprints.keele.ac.uk/5587/1/TattonPhD2018.pdf}.}. We have converted $E(Y-K_{\mathrm{s}})$ to $E(B-V)$ using the extinction coefficients listed in Eqn.~\ref{eqn:ext_coeffs}.

\subsection{Quantitative comparisons}
\label{quantitative_comparisons}

Fig.~\ref{fig:red_maps_comp} presents the literature reddening maps discussed in Section~\ref{literature_reddening_maps} as well as the comparison with the reddening maps
produced in this work based on background galaxies. The top two rows show the literature reddening maps for the outskirts (first row) and bar (second row) regions. The bottom
two rows show the difference between our preferred reddening maps (see Section~\ref{refined_sample_galaxies_preferred_combination_parameters}) and the literature
maps such that $\Delta E(B-V) = E(B-V)_{\mathrm{Bell}} - E(B-V)_{\mathrm{literature}}$. As in Figs.~\ref{fig:red_maps_averoin_gal_passive_stan_aper_refined} and
\ref{fig:red_maps_all}, the contours in the second and fourth rows represent the IRAS 100\,$\mu$m dust emission.

\subsubsection{The outskirts region}
\label{outskirts_region}

Fig.~\ref{fig:red_maps_comp} shows that the reddening maps of the outskirts region of the SMC exhibit similar levels of intrinsic reddening (allowing for the
associated uncertainty in each sample), with median values of $E(B-V)_{\mathrm{med}}=0.02$
and 0.07\,mag for the \cite{Muraveva18} and Tatton et al. (in prep.) samples, respectively. The differential reddening maps imply that, despite sampling the full line-of-sight of
the SMC, the intrinsic reddening values inferred from the background galaxies are typically lower than those inferred from the red clump/RR Lyrae stars. If we allow a factor
of two difference to account for this depth effect, we calculate median differences of $\Delta E(B-V)_{\mathrm{med}}=-0.05$ and $-0.14$\,mag, for the \cite{Muraveva18} and Tatton et al.
(in prep.) samples, respectively. The median uncertainty on the reddening, as derived from the background galaxy sample, is 
$\sigma_{E(B-V)_{\mathrm{med}}}=0.09$\,mag, thus implying that
our reddening map of the outskirts region is entirely consistent with that of \cite{Muraveva18}. Whilst this uncertainty suggests that the reddening map presented here and
that of Tatton et al. are inconsistent, note that there are difficulties in the absolute calibration of the VMC $Y$-band data (see e.g.
\citealp{Gonzalez-Fernandez18,Rubele18}) which could potentially account for the residual differences.

\subsubsection{The bar region}
\label{bar_region}

\begin{landscape}
\begin{figure}
\centering
\includegraphics[width=\linewidth]{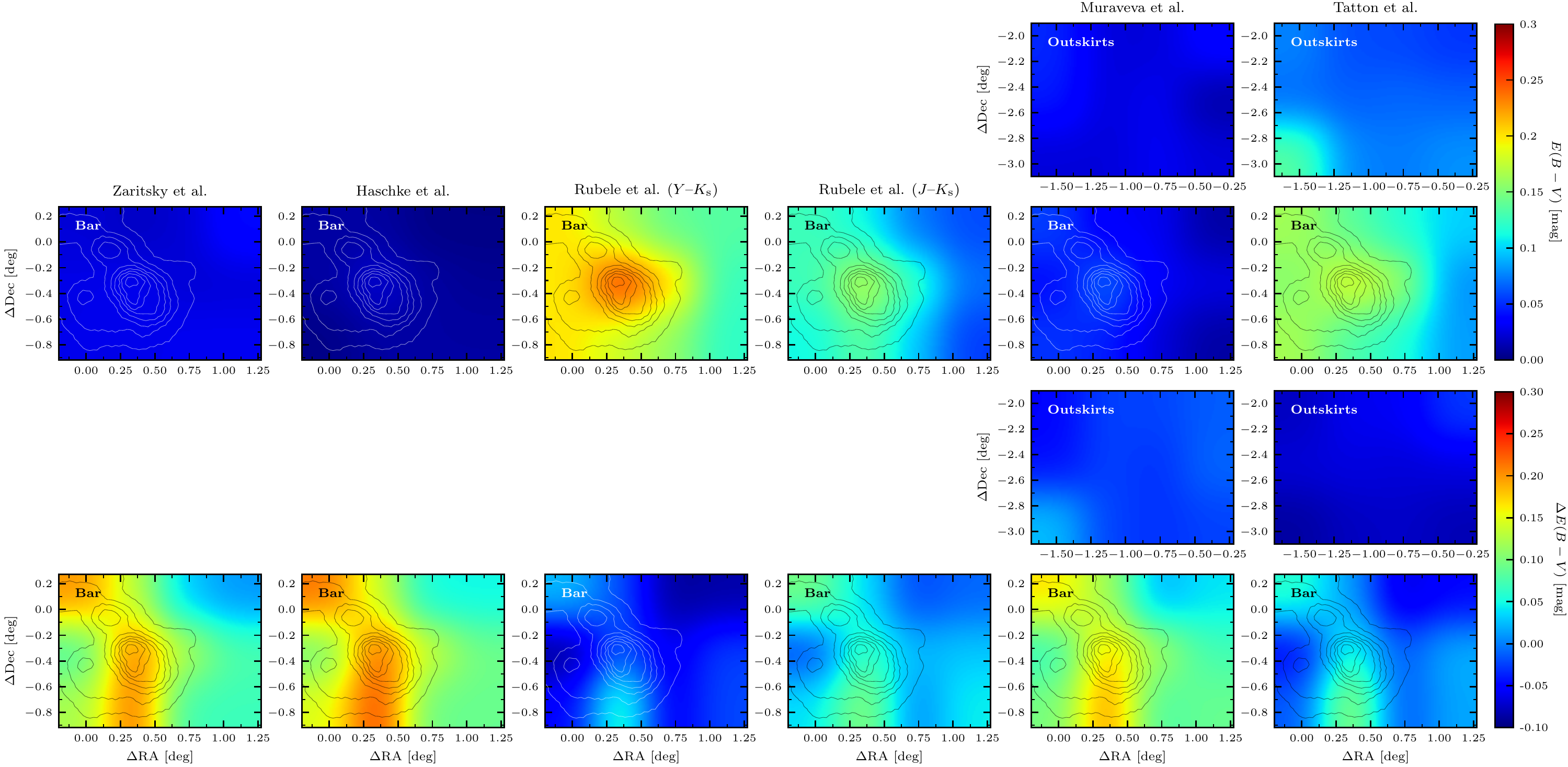}
\caption[]{Gaussian smoothed $20 \times 20$\,arcmin$^{2}$ resolution reddening maps for the outskirts and bar regions of the SMC. Each column denotes a different
literature source. The top two rows show the literature
reddening maps for the outskirts (first row) and bar (second row) regions. The bottom two rows show the difference in $E(B-V)$ between our preferred reddening maps
(see Fig.~\ref{fig:red_maps_averoin_gal_passive_stan_aper_refined}) and the literature maps shown above such that
$\Delta E(B-V) = E(B-V)_{\mathrm{Bell}} - E(B-V)_{\mathrm{literature}}$. The contours in the second and fourth rows represent the IRAS 100\,$\mu$m dust emission.}
\label{fig:red_maps_comp}
\end{figure}
\end{landscape}

From Fig.~\ref{fig:red_maps_comp} it is clear that, even amongst the various literature maps, there is significant variation regarding the intrinsic reddening of the bar
region of the SMC, with the maximum values within a given map ranging from $E(B-V)_{\mathrm{max}}=0.02$\,mag \citep{Haschke11} to
$E(B-V)_{\mathrm{max}}=0.26$\,mag \citep{Rubele18}. Given that the SMC bar is the most productive region, in terms
of star formation (see e.g. \citealp{Cignoni12,Rubele18}), it is interesting to note that several maps, such as those of \cite{Zaritsky02}, \cite{Haschke11} and \cite{Muraveva18},
show no signatures of enhanced intrinsic reddening in this region (although one must make an allowance for the degraded spatial resolution of Fig.~\ref{fig:red_maps_comp}).
Furthermore, whereas one clearly sees a significant increase in the level of intrinsic reddening between the outskirts and bar regions in the Tatton et al. (in prep.) reddening maps,
these two regions are statistically indistinguishable based on the \cite{Muraveva18} reddening values.

These low levels
of reddening could potentially be owing to selection effects, in the sense that the stars used to create these maps lie preferentially in front of the SMC bar and thus do not
include contributions from the extinguished regions associated with recent and ongoing star formation, however this seems rather implausible. In the case of the
\cite{Muraveva18} sample, for which the distances to the RR Lyrae stars are known, even within the bar region the stars span a range of distances ($>$\,10\,kpc) thereby
fully sampling the bar itself. Comparing the
red clump reddening maps of \cite{Haschke11} and Tatton et al. (in prep.), which are essentially derived from the same stars, it is unclear why there is such a systematic
difference when using either optical or near-IR data, as the methods used to determine the reddening of a given star are virtually identical.
Potential systematics could result from differences in the adopted SMC metallicity (which will have an impact on the intrinsic colour adopted to deredden red clump stars),
the different sensitivities of both optical and near-IR wavelengths to interstellar extinction and also potential issues regarding the photometric calibration of the $Y$-band data
(see Section~\ref{outskirts_region}). This latter point may also have an impact on the observed systematic difference between the reddening maps determined from the two sets
of colour-magnitude Hess diagrams in \cite{Rubele18}, where those based on the $K_{\rm{s}}, Y-K_{\rm{s}}$ CMD have noticeably higher reddening values.

Given the amount of variation amongst the various literature maps of the bar region, it is not surprising that the differential reddening maps also vary significantly. Of these,
the most discrepant is the comparison with the red clump reddening map of \cite{Haschke11}, for which we calculate a median difference of
$\Delta E(B-V)_{\mathrm{med}}=0.12$\,mag [maximum difference of $\Delta E(B-V)_{\mathrm{max}}=0.24$\,mag]. Similarly large differences are also seen in the comparison
with the cool star extinction map of \citet[$\Delta E(B-V)_{\mathrm{med}}=0.10$; $\Delta E(B-V)_{\mathrm{max}}=0.22$\,mag]{Zaritsky02} as well as the RR Lyrae reddening
map of \citet[$\Delta E(B-V)_{\mathrm{med}}=0.09$; $\Delta E(B-V)_{\mathrm{max}}=0.20$\,mag]{Muraveva18}. Even if we allow for differences in the depth of the SMC sampled,
it is hard to reconcile residual maximum differences of $\Delta E(B-V)_{\mathrm{max}}=0.23$, 0.19 and 0.16\,mag for the \cite{Haschke11}, \cite{Zaritsky02} and \cite{Muraveva18}
maps, respectively.

In contrast, the two \cite{Rubele18} and the Tatton et al. reddening maps show appreciably smaller differences when compared to our background galaxy
reddening maps [$\Delta E(B-V)_{\mathrm{med}}=-0.04$, 0.04 and 0.00\,mag; $\Delta E(B-V)_{\mathrm{max}}=-0.10$, 0.11 and 0.12\,mag for the $K_{\rm{s}}, Y-K_{\rm{s}}$
and $K_{\rm{s}}, J-K_{\rm{s}}$ \citealp{Rubele18} and Tatton et al. maps, respectively]. Again, allowing for the fact that we sample, on average, twice the depth of the SMC,
we find that both the \cite{Rubele18} $K_{\rm{s}}, J-K_{\rm{s}}$ and Tatton et al. reddening maps are consistent, to within the uncertainties, with our reddening map of the
SMC bar region based on background galaxies. The \cite{Rubele18} $K_{\rm{s}}, Y-K_{\rm{s}}$ reddening map exhibits markedly enhanced levels of reddening
[$\Delta E(B-V)_{\mathrm{med}}=-0.22$\,mag; $\Delta E(B-V)_{\mathrm{max}}=-0.29$\,mag] indicative of a large amount of additional interstellar reddening which our
background galaxies do not suggest. As noted by \citet[see their section~4.1]{Rubele18}, the authors consider the reddening values stemming from the $K_{\rm{s}}, J-K_{\rm{s}}$
CMD analysis more robust, and thus it is reassuring that it is these reddening values which are consistent with those we infer from the background galaxies.

One common feature of all differential reddening maps of the SMC bar region shown in Fig.~\ref{fig:red_maps_comp} is the increased levels of reddening directly below
the bar itself ($\Delta$RA$\,\simeq$\,0.4, $\Delta$Dec$\,\simeq$\,--0.7\,deg) as well as in the north easterly direction along the bar ($\Delta$RA$\,\simeq$\,--0.1,
$\Delta$Dec$\,\simeq$\,0.1\,deg). These enhancements are similar in magnitude to the intrinsic reddening associated with the bar itself and such features are not present
in any of the literature reddening maps in which we see an obvious increase in the intrinsic reddening coincident with the SMC bar. We therefore conclude that these features
in the differential reddening maps stem from our background galaxy reddening maps and it will be interesting to see, when the full analysis of the SMC is complete, whether
additional fluxes for objects close to the edge of a given tile will alleviate such structures.

\section{Summary}
\label{summary}

In this study we have introduced a technique to quantify and map the total intrinsic reddening of two regions within the SMC (the bar and outskirts) based on the analysis
of spectral energy distributions (SEDs) of background galaxies. The main steps involved in achieving this are as follows:

\begin{enumerate}

\item We have used the \textsc{lambdar} photometry package to perform matched aperture photometry on optical and near-IR images taken as part of the SMASH
and VMC surveys, respectively, to create SEDs covering the wavelength range 0.3--2.5$\mu$m. One of the regions is centred on the SMC bar and is
therefore significantly affected by
crowding. Thus, in addition to extracting fluxes using standard (circular) apertures we also adopt PSF-weighted apertures to create two distinct sets of SEDs with which
to determine the intrinsic reddening within the two regions. Photometry was performed on pre-defined sources identified as likely
galaxies from a combination of $K_{s}$, $J$--$K_{\rm{s}}$ colour-magnitude and morphological selections.

\item We adopt the \textsc{lephare} $\chi^{2}$ template-fitting code which allows us to fit the SEDs using a combined set of galaxy/QSO templates. 
We discuss and quantify the effects introduced into the inferred reddening maps as a result of adopting different galaxy templates, apertures and galaxy
samples, as well as several additional criteria designed to refine the sample of galaxies.

\item We find that inferring reddening maps using all available background galaxies and/or QSOs
results in unreliable reddening maps, whereas selecting only galaxies with low levels of intrinsic reddening provides maps which are morphologically consistent with the
large-scale reddening distribution of the SMC as evidenced in previous studies.

\item We find that the use of theoretical galaxy templates results in systematically higher inferred reddening values [by as much as 0.20\,mag in $E(B-V)$] which
is inconsistent with previous determinations of the intrinsic SMC reddening.

\item Finally, we compare our preferred reddening maps (see Section~\ref{refined_sample_galaxies_preferred_combination_parameters})
to stellar reddening maps available in the literature. For the outskirts region we find good agreement between our map and those in the literature (having
taken into consideration that different tracers sample
different lengths along the line-of-sight dust column). The comparison of the bar region, however, is more complicated
owing to the significant variation amongst the various literature reddening maps. For those which clearly demonstrate an enhanced level of reddening associated with the
bar we again find good agreement (accounting for the depth effects). In contrast, for those maps which suggest an extremely low level of intrinsic reddening in the
bar region we find significant discrepancies. Given the lack of any clear cut conclusion in our comparison with the literature maps of the bar region, it is unclear whether this
implies that either our map suggests appreciable levels of dust not accounted for in some stellar reddening maps and/or an issue with our methodology.

\end{enumerate}

In this study we have provided a proof of concept regarding the use of background galaxies observed through the SMC to quantify and map the total intrinsic reddening. Work
is currently underway to extend the techniques introduced herein to the full combined SMASH-VMC footprint which will provide detailed reddening maps covering
over 150\,deg$^{2}$ across the main bodies of both the LMC and SMC, as well as discrete regions within the Magellanic Bridge.

Future large-scale spectroscopic surveys (including several 4MOST consortium surveys; see e.g. \citealp{Cioni19}) will provide tens of thousands of extragalactic spectra
which will provide, not only robust redshift determinations, but also a complementary method to quantify the total intrinsic reddening of the foreground MCs. In this context, and
allowing for the inherent caveats (e.g. uncertain photometric redshifts determinations; see e.g. \citealp*{Salvato19}), our photometrically determined estimates of the dust content
will provide a necessary benchmark to test these against.

\section*{Acknowledgements}

This project has received funding from the European Research Council (ERC) under the European Union's Horizon
2020 research and innovation programme (grant agreement no. 682115).
S.R. acknowledges support from the ERC consolidator grant project STARKEY (grant agreement no. 615604).
Y.C. acknowledges support from NSF grant AST 1655677.
R.R.M. acknowledges partial support from project BASAL AFB-$170002$ as well as FONDECYT project no. 1170364.
The authors would like to thank the Cambridge Astronomy Survey Unit (CASU) and the Wide Field Astronomy
Unit (WFAU) in Edinburgh for providing the necessary data products under the support of the Science and Technology
Facility Council (STFC) in the U.K.
The authors would also like to thank O.~Ilbert for discussions pertaining to the usage of the \textsc{lephare} code as well
as M.~J.~Jarvis for providing the \textsc{lephare} outputs of the combined CFHTLS and VIDEO data sets.
The authors would like to extend their gratitude to the referee whose comprehensive report not only made several excellent
suggestions, but also helped us more clearly present the material.
This project is based on observations collected at the European Organisation for Astronomical Research in the
Southern Hemisphere under ESO programme 179.B-2003.
 Based on
observations at Cerro Tololo Inter-American Observatory,
National Optical Astronomy Observatory (NOAO Prop. ID:
2013A-0411 and 2013B-0440; PI: Nidever), which is operated
by the Association of Universities for Research in Astronomy
(AURA) under a cooperative agreement with the National
Science Foundation
This project used data obtained with the
Dark Energy Camera (DECam), which was constructed by the
Dark Energy Survey (DES) collaboration.
Funding for the DES Projects has been provided by 
the U.S. Department of Energy, 
the U.S. National Science Foundation, 
the Ministry of Science and Education of Spain, 
the STFC of the U.K., 
the Higher Education Funding Council for England, 
the National Center for Supercomputing Applications at the University of Illinois at Urbana-Champaign, 
the Kavli Institute of Cosmological Physics at the University of Chicago, 
the Center for Cosmology and Astro-Particle Physics at the Ohio State University, 
the Mitchell Institute for Fundamental Physics and Astronomy at Texas A\&M University, 
Financiadora de Estudos e Projetos, Funda{\c c}{\~a}o Carlos Chagas Filho de Amparo {\`a} Pesquisa do Estado do Rio de Janeiro, 
Conselho Nacional de Desenvolvimento Cient{\'i}fico e Tecnol{\'o}gico and the Minist{\'e}rio da Ci{\^e}ncia, Tecnologia e Inovac{\~a}o, 
the Deutsche Forschungsgemeinschaft, 
and the Collaborating Institutions in the Dark Energy Survey. 
The Collaborating Institutions are 
Argonne National Laboratory, 
the University of California at Santa Cruz, 
the University of Cambridge, 
Centro de Investigaciones En{\'e}rgeticas, Medioambientales y Tecnol{\'o}gicas-Madrid, 
the University of Chicago, 
University College London, 
the DES-Brazil Consortium, 
the University of Edinburgh, 
the Eidgen{\"o}ssische Technische Hoch\-schule (ETH) Z{\"u}rich, 
Fermi National Accelerator Laboratory, 
the University of Illinois at Urbana-Champaign, 
the Institut de Ci{\`e}ncies de l'Espai (IEEC/CSIC), 
the Institut de F{\'i}sica d'Altes Energies, 
Lawrence Berkeley National Laboratory, 
the Ludwig-Maximilians Universit{\"a}t M{\"u}nchen and the associated Excellence Cluster Universe, 
the University of Michigan, 
{the} National Optical Astronomy Observatory, 
the University of Nottingham, 
the Ohio State University, 
the University of Pennsylvania, 
the University of Portsmouth, 
SLAC National Accelerator Laboratory, 
Stanford University, 
the University of Sussex, 
and Texas A\&M University.
Finally, this project has made extensive use of the Tool for OPerations on Catalogues And
Tables (TOPCAT) software package \citep{Taylor05} as well as the following open-source
\texttt{Python} packages: Astropy \citep{Astropy18}, matplotlib \citep{Hunter07}, NumPy
\citep{Oliphant15}, pandas \citep{McKinney10}, and SciPy \citep{Jones01}.

\bibliographystyle{mn3e}
\bibliography{references}

\appendix

\section{Refining our sample of galaxies}
\label{refining_sample_galaxies}

In this Appendix we discuss the three criteria used to refine our sample of galaxies from which the intrinsic reddening maps
of the bar and outskirts of the SMC are inferred.

\subsection{Number of bandpasses}
\label{number_of_bandpasses}

For the vast majority of objects across all samples, the SEDs comprise the full eight bandpasses available from our combined optical/near-IR data set
(see Table~\ref{tab:catalogue_properties}). Thus, the simplest constraint to ensure the most robust SED fits is to remove objects for which the SEDs have
fewer than eight bandpasses.

\subsection{Blending and incompleteness at faint magnitudes}
\label{blending_incompleteness}

\begin{figure}
\centering
\includegraphics[width=\columnwidth]{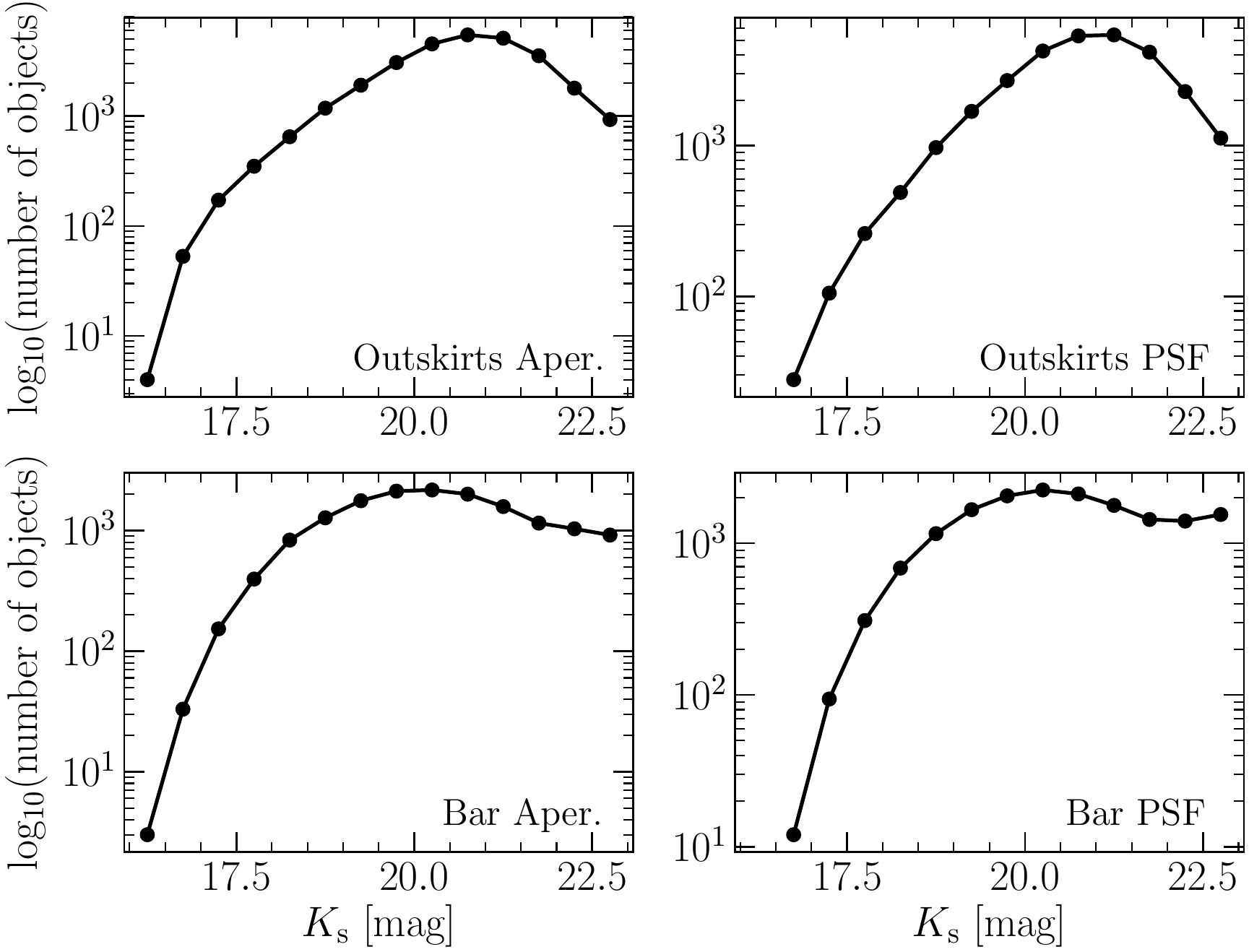}
\caption[]{$K_{\rm{s}}$-band AB magnitude differential number counts at 0.5\,mag intervals for objects in both the standard and PSF-weighted aperture samples of
the bar and outskirts regions of the SMC. Incompleteness starts to become an issue in the bar region approximately 1\,mag brighter than in the outskirts region,
which is presumably due to the combined effects of crowding and reddening.}
\label{fig:ks_comp}
\end{figure}

From Fig.~\ref{fig:cmd_sharp} it is clear that as one moves to fainter magnitudes, the reliability
with which we are able to differentiate between stellar and non-stellar objects deteriorates. Although we have attempted to remove potential contaminants
using a combination of colour-magnitude and morphological selections, from Fig.~\ref{fig:cmd_sharp} it is evident that at an apparent magnitude of
$K_{\rm{s}}$\,$\simeq$\,20\,mag there is an increase in the stellar probability of the objects within our initial colour-magnitude selection. Furthermore, although
we have introduced a cut on the stellar probability of less than 0.33, there is sufficient ambiguity regarding the robustness of these values
(based on the criteria used to assign these probabilities; see Section~\ref{identification_of_galaxies}) in this region of colour-magnitude space. Finally, Fig.~\ref{fig:cmd_sharp}
also shows that our $K_{\rm{s}}$-band sharpness index cut of Sharp$_{K_{\rm{s}}} > 0.5$ is probably only sufficient to reliably distinguish between
stellar and non-stellar objects for apparent magnitudes brighter than $K_{\rm{s}}$\,$\simeq$\,20\,mag. At fainter magnitudes the $K_{\rm{s}}$-band sharpness index
distribution blooms such that both stellar and non-stellar objects can have values indicative of extended objects. Note that the CMD shown in Fig.~\ref{fig:cmd_sharp}
is in the natural VISTA system whereas the magnitudes we have calculated using \textsc{lambdar} are AB magnitudes (see Section~\ref{lambdar_photometry} for details), and so we must
account for this difference when defining a magnitude cut\footnote{$K_{\mathrm{s,AB}} = K_{\mathrm{s,VISTA}} + 1.838$\,mag \citep{Gonzalez-Fernandez18}.}.

We must also consider the effects of incompleteness in our samples and how this may vary
due to both environmental effects such as crowding and reddening as well as the adopted apertures from which the fluxes are calculated.
Fig.~\ref{fig:ks_comp} shows the $K_{\rm{s}}$-band AB magnitude differential number counts at 0.5\,mag intervals for objects in both the standard and PSF-weighted aperture
samples of the bar and outskirts regions of the SMC. From Fig.~\ref{fig:ks_comp} we can see that incompleteness starts to become an issue at a $K_{\rm{s}}$-band AB
magnitude of $\simeq$\,20\,mag in both outskirts region samples, whereas in the bar region samples it is approximately 1\,mag fainter. To account for this, we
adopt apparent $K_{\rm{s}}$-band AB magnitude cuts of 21.0 and 20.5\,mag for the standard aperture samples and 21.5 and 20.5\,mag for the
PSF-weighted aperture samples of the outskirts and bar regions, respectively.

\subsection{Redshift probability distributions with multiple peaks}
\label{multiple_peaked_pdfz}

\begin{figure}
\centering
\includegraphics[width=\columnwidth]{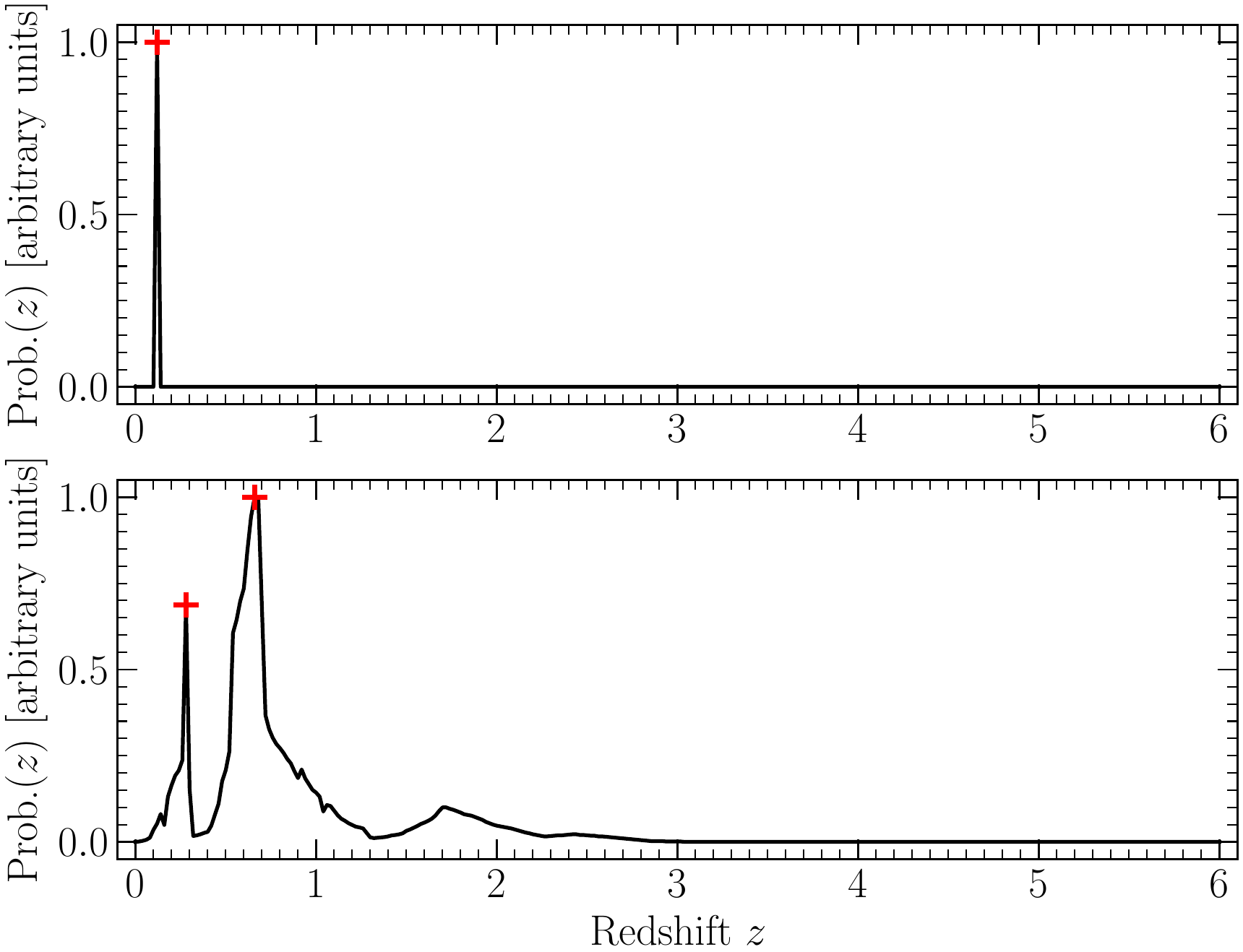}
\caption[]{Redshift probability distribution functions (PDF$z$s) of two objects in the outskirts PSF-weighted aperture sample. \emph{Top:} A unimodal distribution
characterised by a single peak with a reliable redshift measurement at $z=0.12$. \emph{Bottom:} A multimodal distribution characterised by multiple peaks. The highest
probability redshift is at $z=0.66$, however there is a strong secondary peak at $z=0.28$ as well as an additional lower probability peak at $z=1.71$. The
red crosses in both panels denote the significant peaks identified within the PDF$z$s used to identify objects with multimodal PDF$z$s (see text for details).}
\label{fig:pdfz_uni_multi}
\end{figure}

\textsc{lephare} automatically creates a redshift probability distribution function (PDF$z$) for each object thus
providing a metric to study the redshift probability for a given object over a given redshift range (in this case 0\,$\leq$\,$z$\,$\leq$\,6). Fig.~\ref{fig:pdfz_uni_multi} demonstrates the,
broadly speaking, two types of PDF$z$s; a single-peaked (unimodal) distribution with a prominent peak at a given $z$ and very little dispersion and a multi-peaked (multimodal) distribution
with numerous peaks (sometimes at comparable probabilities) and a high level of dispersion. Although both PDF$z$s will provide a best-fitting
photometric redshift, it is clear that both of these are not equally reliable. Given that we have the added complication of including reddening in the SED fitting
process, it would make sense to ensure that the sample of objects from which we investigate the reddening properties comprise those with the most robust redshift determinations,
thereby minimising the degeneracy between
redshift and reddening. Although \textsc{lephare} includes an inbuilt function to detect secondary peaks above a given threshold in the output PDF$z$s, it is unclear
what this threshold is and, interestingly, the PDF$z$ shown in the bottom panel of Fig.~\ref{fig:pdfz_uni_multi} is not flagged as an object for which the PDF$z$ contains a
secondary peak. We therefore decide to screen the PDF$z$s of each object to identify multimodal distributions using the \texttt{Python} routine \texttt{scipy.signal.find\_peaks}.
For the various input parameters, we adopt a value of 0.3 for the minimum peak height (the maximum probability in the PDF$z$s is set to unity), a value of 0.1 in $z$ space
for the minimum peak distance, and a threshold of zero (given that some of our unimodal distributions are broader than that shown in the top panel of Fig.~\ref{fig:pdfz_uni_multi},
but also harbour no additional peaks in the PDF$z$).
We therefore retain those objects for which the former two constraints are met in addition to those objects for which we detect only one peak in the corresponding PDF$z$.

\end{document}